\documentclass[final]{elsarticle}
 \usepackage{geometry}
\usepackage{subcaption}

\usepackage[utf8]{inputenc}

\usepackage{multicol}
\usepackage{enumitem}   

\usepackage{amsfonts}
\usepackage[version=3]{mhchem}
\usepackage{mathtools}
\usepackage{cases}
\usepackage[overload]{empheq}
\usepackage{bbm}
\usepackage{amsthm}
\usepackage{caption}
\usepackage{subcaption}
\usepackage{algorithm}
\usepackage{algpseudocode}
\usepackage{pdfpages}
\usepackage{natbib}
\usepackage{graphicx}
\usepackage{amsmath}
\usepackage{dsfont}
\usepackage{float}
\usepackage{xcolor}
\usepackage{geometry}
\newtheorem{theorem}{Theorem}[section]

\newtheorem{lemma}[theorem]{Lemma}
\newtheorem{assumption}{Assumption}[section]
\newtheorem*{remark}{Remark}

\begin{document}

\begin{frontmatter}
\title{Safe model-based design of experiments using Gaussian processes}
\author[a]{P. Petsagkourakis}
\author[a]{F. Galvanin\corref{cor2}}
 \ead{f.galvanin@ucl.ac.uk}
 \cortext[cor2]{Corresponding author}
 \address[a]{Centre for Process Systems Engineering (CPSE), Department of Chemical Engineering, University College London, Torrington Place, London WC1E 7JE, United Kingdom}

\begin{abstract}
The construction of kinetic models has become an indispensable step in developing and scale-up of processes in the industry. Model-based design of experiments (MBDoE) has been widely used to improve parameter precision in nonlinear dynamic systems. Such a framework needs to account for both parametric and structural uncertainty, as the physical or safety constraints imposed on the system may well turn out to be violated, leading to unsafe experimental conditions when an optimally designed experiment is performed. In this work, Gaussian processes are utilized in a two-fold manner: 1) to quantify the uncertainty realization of the physical system and calculate the plant-model mismatch, 2) to compute the optimal experimental design while accounting for the parametric uncertainty. TheOur proposed method, Gaussian process-based MBDoE (GP-MBDoE), guarantees the probabilistic satisfaction of the constraints in the context of the model-based design of experiments. GP-MBDoE is assisted with the use of adaptive trust regions to facilitate a satisfactory local approximation. The proposed method can allow the design of optimal experiments starting from limited preliminary knowledge of the parameter set, leading to a safe exploration of the parameter space. This method's performance is demonstrated through illustrative case studies regarding the parameter identification of kinetic models in flow reactors.
\end{abstract}

\end{frontmatter}
\section{Introduction}
Biochemical, physicochemical or catalytic processes are often too complex to develop accurate physics-based only models, and plant-model mismatch is inevitable.
It is common to develop approximate kinetic models under some assumptions regarding the system's mechanism to represent the physical system accurately. The availability of a trustworthy kinetic model could be used to predict the system's behaviour outside of the validated experimental conditions and then be employed for design, optimization and control in process systems engineering applications \cite{Bonvin2016}. 
The model identification requires experimental data to evaluate the validity of a proposed model and estimate the model parameters to match the physical system's behaviour over the selected range of operating conditions. Nevertheless, the experimental procedure may become a costly, very time consuming and infeasible task when poor experimental designs are implemented. 
Identifying an appropriate approximated model is highly dependent on limitations due to the certain observability of the physical phenomena. Additionally, the respected data may be difficult and expensive to obtain because of physical accessibility and/or limitations on the experimental budget.

Model-based design of experiments (MBDoE) has been widely utilized to improve parameter precision in highly nonlinear dynamic systems. 
The main goal of MBDoE is to design control inputs and sampling schedules \cite{Franceschini2008} to achieve the most informative experiment possible by satisfying constraints on feasibility and safety of operation.
The optimal MBDoE problem for improving parameter estimation involves the maximization of a measure of the expected information (or the minimization of a measure of the expected variance of model parameters) by acting on experiment decision variables to ensure the feasibility of the experimental trials. However, since the optimization problem is based on the available model, both plant-model mismatch and parametric uncertainty may significantly affect the result \cite{PETSAGKOURAKIS2020106649, pan2020constrained}.

The literature consists of various approaches to ensure optimality and feasibility, particularly under the presence of parametric uncertainty \cite{PRONZATO1985103}. Two distinctly different scenarios are considered regarding the information and the constraint satisfaction, robust and stochastic approaches. A robust approach for the information content was proposed \cite{Asprey2002}, where the predicted information content of experiments needs to be optimal for the whole space of the uncertain parameters.
The robust optimization problem is usually solved in terms of min-max optimization, in which uncertainties are typically assumed to be deterministic and bounded \cite{PRONZATO1985103, Korkel2004327,WELSH201113197,ROJAS2007993}. This problem is often numerically intractable when candidate models are highly nonlinear. Various methods have been proposed to accommodate this limitation, including linearization of the inner optimization \cite{Korkel2004327} and a sensitivity-based approximate robust approach assuming ellipsoidal joint confidence region model parameters \cite{Nimmegeers2020}. The robust approaches are typically conservative as they require the optimality and satisfaction of constraints to hold for all the possible values within the predefined uncertainty region. 
On the contrary, a stochastic approach has been popularized, where the information content is maximized in expectation, and the constraints are satisfied with a given probability \cite{Asprey2002}. 
The stochastic (also called probabilistic) experimental design framework avoids the typical strong conservatism of the worst-case MBDoE techniques, as the probability of occurrence of different realizations is accounted for differently. Various approaches \cite{Telen2014, Nimmegeers2020, Galvanin_backoff, Streif2014, Mesbah2015} have been proposed in this family of problems usually inspired by the techniques originated from optimal control problem.

The satisfaction of chance constraints has been widely studied in the field of optimal control, especially from the stochastic model predictive control community (SMPC). Comprehensive reviews regarding SMPC can be found in \cite{Mesbah2016, Farina2016, Geletu2013}. Additionally, both parametric and structural mismatches have been incorporated in \cite{Arcari2019,  Hewing2020, Arellano-Garcia2020}. Probabilistic (chance) constraints are incorporated into the MBDoE problem to seek a trade-off between a designed experiment's information content and allow for a user-defined level of risk during the experiment. Most approaches compute the expected information given the (previous) parameter identification procedure, where confidence regions are constructed using approximation techniques \cite{Peric2018} around the point estimates for the parameters. 

Galvanin et al. \cite{Galvanin_backoff} directly utilized a Monte Carlo approach to compute the constraint tightening (backoff) given the approximated probability distribution of the parameters. Such a method is very accurate for estimating the constraint tightening for the user-defined constraint satisfaction but computationally costly.
Telen et al.\cite{Telen2014} proposed the use of the unscented transformation \cite{VanDerMerwe2001} to propagate the parametric uncertainty in the objective function and the constraints. The unscented transformation has widely be utilized in the optimal control \cite{Bradford2018a} due to the low computational complexity. Nevertheless, unscented transformation approaches may fail in the presence of strong nonlinear relationships between the parameters and the states of the model. 
In this context, polynomial chaos expansion (PCE) has been used to compute the required moments of the objective and constraints in \cite{Mesbah2015} and recently in \cite{Nimmegeers2020}. Polynomial chaos expansions have been used in stochastic model predictive control for the purposes of uncertainty propagation through a nonlinear model \cite{Wiener}. Similarly, Gaussian process (GP) \cite{Rasmussen2006, Wang2020} has been proposed as an alternative to PCE, as they are non-parametric
models. Gaussian processes have been employed due to the estimation of uncertainty of prediction around points caused by an insufficient amount of data-points. The GP usually assumes an independent and identically distributed (iid) Gaussian noise added in the observed measurements, and the outputs obtained at different inputs are jointly Gaussian distributed, where a kernel function defines the covariance. Gaussian process has not found a significant application in  MBDoE, with exception in \cite{Olofsson2018} in the context of model discrimination. In this case the GP is used to approximate the different models and marginalize the parameters of the models given the computed uncertainty. 
Nevertheless, GP has extensively be used in the area of optimal control, where the prior models and black-box techniques are combined to describe the system. GP may be coupled with approximate method for propagating the uncertainty as unscented Kalman Filter \cite{Ko2007}, exact moment matching \cite{Girard2002,Deisenroth2015}, linearization \cite{Girard2002} and scenario-based approaches \cite{Umlauft2018,BRADFORD2020106844}. Recently, GPs have been used in hybrid modeling rational, where they are coupled with the prior (nominal) model that is assumed for the system, and learning is conducted only for the GP \cite{del2020modifier, Hewing}.

The methods of robust MBDoE, e.g. \cite{Asprey2002, Nimmegeers2020} use the confidence regions computed in the parameter estimation step. Such an assumption may be sufficient to accommodate the uncertainty for the objective of the information content. However, constraint satisfaction should require different treatment. The satisfaction of the constraints is typically guaranteed given the uncertainty of the parameters and
the required assumptions, including the availability of a large number of available prior experimental data points. In the typical scenario of paucity of data at the beginning and a presence of model mismatch, these assumptions do not hold. 
As the methodology is model-based, both model mismatch (i.e., a model structure inadequate to represent the physical systems ) and parametric mismatch (i.e., incorrect values of the parameters ) may affect the consistency of the whole design procedure \cite{Peric2018, petsagkourakis2020constrained}.
These assumptions have been avoided in some works using a disturbance based estimation \cite{Galvanin2012_DIST} and the MBDoE in the context of guaranteed parameter estimation \cite{Reddy2017}. Additionally, Quaglio et al.\cite{Quaglio2018b} developed a method to find the region where the approximated model is valid and avoid the intrinsic issues with the model-mismatch in MBDoE methods \cite{QUAGLIO2018515}.

In this work, we propose a novel MBDoE framework suitable for guaranteeing operation feasibility when a plant-model mismatch is strongly present, i.e. possible disturbances/hardware malfunctions and the initial limited data. The plant-model mismatch is a critical issue in optimal control, where stability\cite{automatica_stability, Petsagkourakis_TAC}, offset-free tracking \cite{Paulson2019} and satisfaction of constraints has been extensively studied \cite{Hewing}.

We treat the objective function of the MBDoE by utilizing a stochastic surrogate, i.e. GP. Additionally, we focus on the satisfaction of constraints that has not been addressed in the literature for the optimal experimental designs avoiding the use of covariance of the parameters that have been computed from the parameter estimation. Gaussian process is utilized to compute the model mismatch together with a trust region around operational point, that is updated iteratively. This is inspired by the derivative-free methods \cite{Conn2005} that were recently applied in  \cite{del2020modifier}. This way, restrictive assumptions of the nature of the `real' model are avoided.

The structure of this paper is as follows. The problem definition is outlined in section 2, the details of the proposed method for safe, optimal experimental design is presented in section 3. A case study is presented in section 4, where the framework is applied to two separate case studies, and in the last section, conclusions are outlined.

\section{Problem Definition}
We assume that the process under consideration can be represented by a set of differential and algebraic equations (DAEs), where measured variables $\textbf{y}$ can only be measured in finite sampling times $\{1,\dots,N\}$
\begin{equation}\label{eq:real}
    \begin{split}
        &\textbf{f}(\dot{\textbf{x}}, \textbf{x}, \textbf{u},\textbf{w}, {\pmb{\theta}}^*) = 0\\
        &\textbf{y} = \textbf{h}(\textbf{x},\textbf{u},\pmb{\nu}),
    \end{split}
\end{equation}
where $\textbf{x}\in \mathbb{R}^{n_x}$, $\textbf{u}\in\mathbb{R}^{n_u}$, $\textbf{y}\in\mathbb{R}^{n_y}$ and ${\pmb{\theta}}^*\in\mathbb{R}^{n_{\theta^*}}$ are the vector of state variables representing the true system, vector of control (manipulated) variables, vector of measured (output) variables and parameters of the process, correspondingly. Additionally, the physical system is assumed to be  subject to additive normally distributed disturbance ($\textbf{w}\in\mathbb{R}^{n_w}$) and noise ($\pmb{\nu}\in\mathbb{R}^{n_\nu}$). The functions $\textbf{f}: \mathbb{R}^{n_x}\times\mathbb{R}^{n_u}\times  \mathbb{R}^{n_w}\rightarrow \mathbb{R}^{n_x}$ and $\textbf{h}:\mathbb{R}^{n_x}\times\mathbb{R}^{n_u}\times  \mathbb{R}^{n_w}\rightarrow \mathbb{R}^{n_y}$ are unknown functions that describe the observed process. For practical reasons this model will be refereed as `true model'.

In practice, the mapping relating the process inputs and outputs of (\ref{eq:real}) is typically unknown, and only an approximate model is available, parametrized by a set of parameters $\pmb{\theta}$. However, in many cases \cite{PETSAGKOURAKIS2020106649, arcari2019dual} it is unlikely that a model's predictions coincide with those of the real system due to the relevant assumption, like the possible plant-model mismatch and/or equipment malfunction (e.g. bias in the pump, wrongly calibrated HPLC). Let the approximated model derived from simplifying hypotheses and/or ignoring equipment's malfunctions be: 
\begin{equation}\label{eq:model}
    \begin{split}
        &\hat{\textbf{f}}(\dot{\hat{\textbf{x}}}, \hat{\textbf{x}}, \textbf{u},\pmb{\theta}) = 0\\
        &\hat{\textbf{y}} = \hat{\textbf{h}}(\hat{\textbf{x}},\textbf{u}),
    \end{split}
\end{equation}
where $\hat{(\cdot)}$ represents the model predictions. The expected values of the  parameters $\pmb{\theta}$ and their corresponding distribution are estimated given available measurements. Notice that the vector of state variables for the model ($\hat{\textbf{x}}$) is not necessarily the one from (1), as different models may have different states to describe the behaviour. 

Optimal model-based design of experiments (MBDoE) for parameter estimation is employed to reduce the variance of the parameters $\pmb{\theta}$, which is part of the model identification procedure. This technique is usually embodied in a sequence of three key activities: experimental design (MBDoE), experimental execution and parameter estimation. Such framework is particularly flexible, allowing for the definition of a set of active constraints on both state and design variables during the optimization.  At the experimental design step, MBDoE is usually inferred as an optimization problem in which the optimal experimental conditions are determined (control variables $\textbf{u}$, sampling times, duration of experiments) to minimize a measure of the predicted covariance matrix or maximize a measure of the expected Fisher information matrix (FIM). Most importantly, the designed experiments should take into account feasibility specification (e.g. safety constraints) during the experimental trials.
\subsection{Maximum Likelihood Estimation \& Confidence Regions}\label{sec:mle}
The uncertainty of the model parameter is computed using the maximum likelihood estimation results, and the respective confidence regions are used for the robust/stochastic optimally informative experiment or robust/chance constraint satisfaction.
In this subsection, the most common methods to construct confidence regions are discussed, which are usually used for addressing the feasibility problem of chance constraints. The limitation of such approaches is also given.

Let $n_{b}$ be the experiments with $n_m$ measurements be available $\mathcal{D}=\{\textbf{y}_b(t_m),\textbf{u}_b(t_m),\textbf{x}_m(0) , \forall b, m\in n_b, n_m\}$. If the manipulated variables are assumed to be precisely known, then the joint probability of the prediction-observation mismatch in the available measurements for the parameter values $\pmb{\theta}$ is the following log-likelihood:
\begin{equation}\label{eq:loglike}
\begin{split}
-2\log\left(\mathcal{L}(\pmb{\theta}|\mathcal{D})\right)&=\log(2\pi^{ n_bn_m})+\sum_{b=1}^{n_b}\sum_{m=1}^{n_{meas}}\log\left(|\pmb{\Sigma}_{mi}^y|\right)\\
&+ \left( {\hat{\textbf{y}}_{b}(t_m,\pmb{\theta})-\textbf{y}_{b}(t_m)}\right)^T \pmb{\Sigma}^{y}_{bm}\left( {\hat{\textbf{y}}_{b}(t_m,\pmb{\theta})-\textbf{y}_{b}(t_m)}\right),
\end{split}
\end{equation}
where $\hat{\textbf{y}}_{b}(t_m,\pmb{\theta})$ corresponds to the solution of (\ref{eq:model}). The maximum-likelihood estimation computes  $\hat{\pmb{\theta}}$ to maximize the likelihood function ($\mathcal{L}$) or minimize ($\ref{eq:loglike}$) subject to (\ref{eq:model}).
Confidence regions can be computed, under the assumption that
model (\ref{eq:model}) is correct and the respected computed parameters parameters ($\hat{\pmb{\theta}}$) are close to the (unique) ‘true’
values of the model parameters ($\hat{\pmb{\theta}}$). Then for a large number of available measurements, both the likelihood subset ratio statistic and the Wald subset statistic follows a chi-square distribution \cite{Seber2015, Meeker199548}. These asymptotic confidence results can be used to obtain  $100(1-\alpha$)$\%$  likelihood-based, $\Theta_\mathcal{L}$, (\ref{eq:likelihood_stat}) and $\Theta_\mathcal{W}$ Wald~(\ref{eq:wald_stat}) confidence regions:
\begin{equation}\label{eq:likelihood_stat}
    \Theta_\mathcal{L} :=-2\text{log}\left[\dfrac{\mathcal{L}({\pmb{\theta}}|\mathcal{D})}{\mathcal{L}(\hat{\pmb{\theta}}|\mathcal{D})}\right]\leq \chi^2_{n_{dof}}(1-\alpha),
\end{equation} 
\begin{equation}\label{eq:wald_stat}
    \Theta_\mathcal{W} :=(\pmb{\theta} - \hat{\pmb{\theta}})^T \textbf{V}_\theta (\pmb{\theta} - \hat{\pmb{\theta}})\leq \chi^2_{n_{dof}}(1-\alpha),
\end{equation}
with $\textbf{V}_\theta$ being the covariance for the parameters calculated at $\hat{\theta}$ and  $\chi^2_{n_{dof}}(1-\alpha)$ is the $(1-\alpha)$ quantile of the chi-square distribution with $n_{dof}$ degrees of freedom. The covariance $\textbf{V}_\theta$ is usually approximated by a Taylor expansion of the likelihood. Particularly,   (\ref{eq:wald_stat}) has  extensively been utilized in the design of experiments \cite{Nimmegeers2020, Asprey2002} as the closed-form expression is available compare to the likelihood-subset ratio~(\ref{eq:likelihood_stat}), where there is no closed-form expression available \cite{doi:10.1002/aic.690470811}. Nevertheless, the Wald confidence regions depend on the model reparameterization because of the (approximate) covariance term $\textbf{V}_\theta$, which may significantly affect  the resulted confidence regions \cite{Quaglio_reparam}. 
Additionally, in the same rationale, a Laplace approximation can be used for the posterior distribution of the parameter estimates given the available data:
\begin{equation}\label{eq:Laplace}
    \begin{split}
        p(\pmb{\theta}|\mathcal{D}) \sim \mathcal{N}(\pmb{\hat{\theta}},\textbf{V}_\theta),
    \end{split}
\end{equation}
where  $\textbf{V}_\theta$ is the covariance for the parameters at $\hat{\theta}$, similar to (\ref{eq:wald_stat}). The parametric uncertainty estimation of (\ref{eq:likelihood_stat}-\ref{eq:Laplace}) are the most common used strategies to approximate confidence regions or posterior distribution of the parameters. Two additional methods can be used that are more computationally demanding: Bayesian-based estimation \cite{Hastings} and set-membership estimation \cite{Peric2018}. 

In the context of stochastic MBDoE, these strategies are used to exploit (\ref{eq:likelihood_stat}-\ref{eq:Laplace}) to formulate a stochastic optimization problem subject to chance constraints. The parametric uncertainty estimation given by (\ref{eq:likelihood_stat}-\ref{eq:Laplace}) is affected by several limitations:  \\
i) A large amount of data is required. However, this is not the case when we design experiments since we want to keep the number of data-points to the minimum.\\
ii) The covariance matrix $\textbf{V}_\theta$ is approximated by a Taylor expansion. This means that the highly nonlinear models will result in unreliable covariance functions. \\
iii) The expression in (\ref{eq:likelihood_stat}) is more accurate but with disadvantages (see point (i)) and there is no closed-form expression. Hence, Monte Carlo simulations or other numerical solutions \cite{doi:10.1002/aic.690470811} are required.

\subsection{General Problem Formulation}
The general formulation of the problem is constructed here where 1) the constraints are satisfied for a true model, and 2) the new experiment is the most informative for a given approximated model. The following infinite-dimensional problem can describe the proposed experimental design, $\mathcal{P}(\textbf{u})$ with $\textbf{u}$ being the optimization variables:
        \begin{align}[left = {\mathcal{P}(\textbf{u}):=}\enspace\empheqlbrace]
           &\begin{alignedat}{2}\label{eqobj_inf}
 &\max_{\textbf{u}}\mathbb{E}_{\pmb{\theta}\sim \Theta}\left(\psi\left(\dfrac{\partial \hat{\textbf{x}}(t_N)}{\partial \pmb{\theta}}^\top\Sigma_{exp}\dfrac{\partial \hat{\textbf{x}}(t_N)}{\partial \pmb{\theta}}+\textbf{M}_0\right)\right)\\
    &\text{s.t.}\end{alignedat}\\
            &   \begin{alignedat}{2} \label{eq:approx_inf}
          &  \text{Approximated Model: } & \\&     \hat{\textbf{x}}(t_0) = \textbf{x}(0)\\
   &\hat{\textbf{f}}(\dot{\hat{\textbf{x}}}, \hat{\textbf{x}}, \textbf{u},\pmb{\theta}) = 0\\
        &\hat{\textbf{y}} = \hat{\textbf{h}}(\hat{\textbf{x}},\textbf{u})
         \end{alignedat}\\
                     &   \begin{alignedat}{2} \label{eq:real_inf}
          &  \text{Real System: } & \\&          \textbf{x}(t_0) = \textbf{x}(0)\\
 &\textbf{f}(\dot{\textbf{x}}, \textbf{x}, \textbf{u},\textbf{w}) = 0\\
       & \textbf{y} = \textbf{h}(\textbf{x},\textbf{u},\pmb{\nu})         \end{alignedat}\\
      & \begin{alignedat}{2} \label{eq:constr_inf}
          &  \text{Constraints: } & \\& \textbf{u}\in\mathbb{U}\\
    &\mathbb{P}(g_j\left(\textbf{x},\textbf{u} \right)\in \mathbb{X})\geq 1-\bar{\alpha}, \forall j \in \{1,\dots, n_g\},
         \end{alignedat}
        \end{align}
where $\psi(\cdot)$ is a scalar function of the Fisher information (it depends on the selected design criterion, e.g. A-optimal, D-optimal etc), that is maximized in expectation given ${\Theta}$ and $\textbf{M}_0$ is the prior Fisher information matrix representing the prior knowledge on the parametric system as derived from preliminary experiments. Additionally, the equality path constraints (\ref{eq:approx_inf}) corresponds to the DAE that describes the approximated model parametrized by $\pmb{\theta}$ and characterized by states $\hat{\textbf{x}}$, the chance constraints (\ref{eq:constr_inf}) are subject to the real behaviour of the physical system that in this case is represented as a stochastic dynamic system (\ref{eq:real_inf}) with states $\textbf{x}$. Additionally, the constraints are meant to be satisfied with probability higher than  $1- \bar{\alpha}\in \left[0,1\right]$, a user-defined parameter representing the probability of constraint satisfaction with values close to 1.
\subsection{Standard Framework}
In this section the standard framework is described. First, each stage of the standard framework for MBDoE is briefly given.

\begin{figure}[H]
\begin{subfigure}[t]{0.5\textwidth}
  \centering
  \includegraphics[width=.97\linewidth]{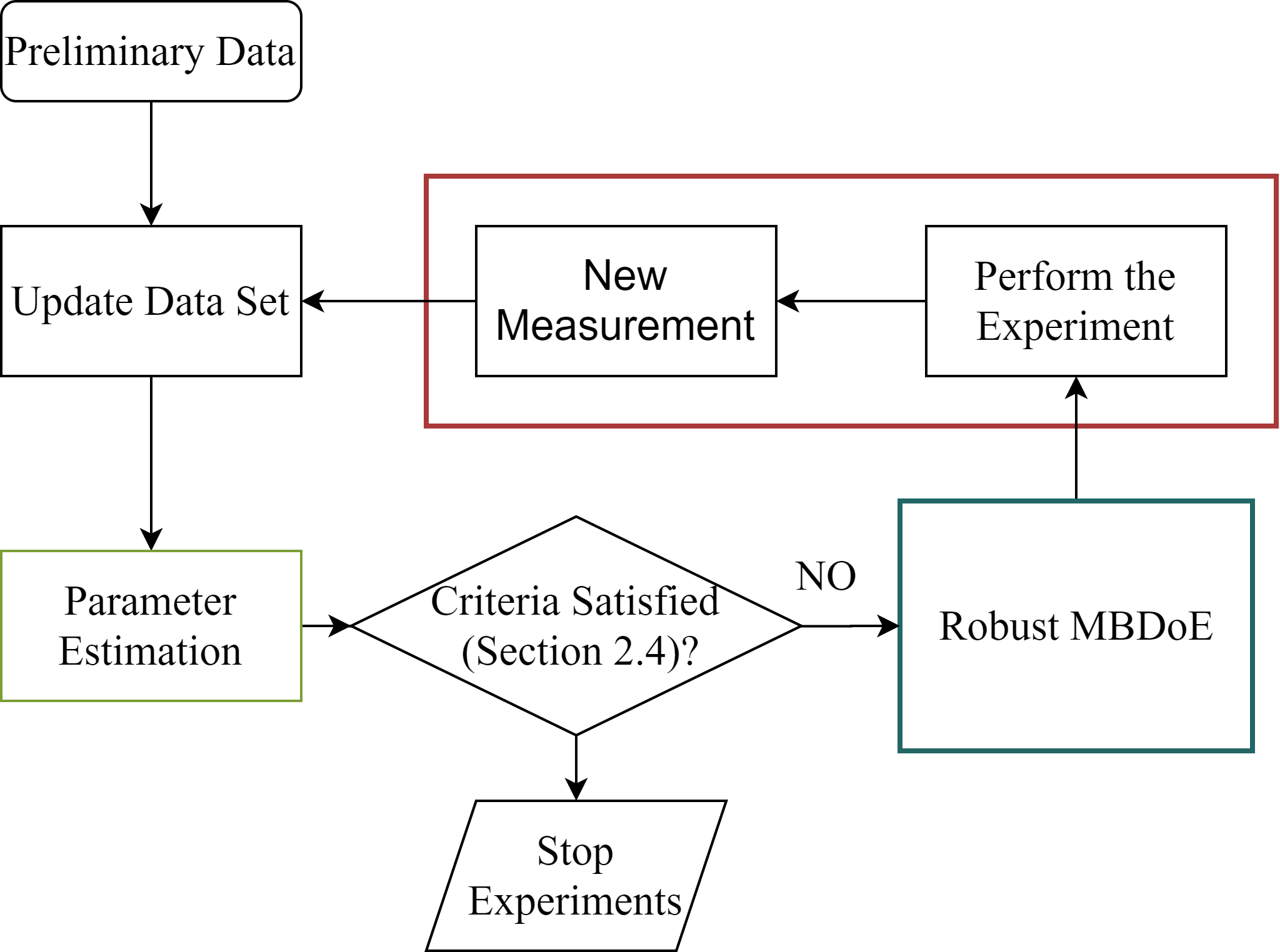}
  \caption{Standard framework for MBDoE}
  \label{fig:overall_a}
\end{subfigure}
\begin{subfigure}[t]{0.5\textwidth}
  \centering
  \includegraphics[width=.98\linewidth]{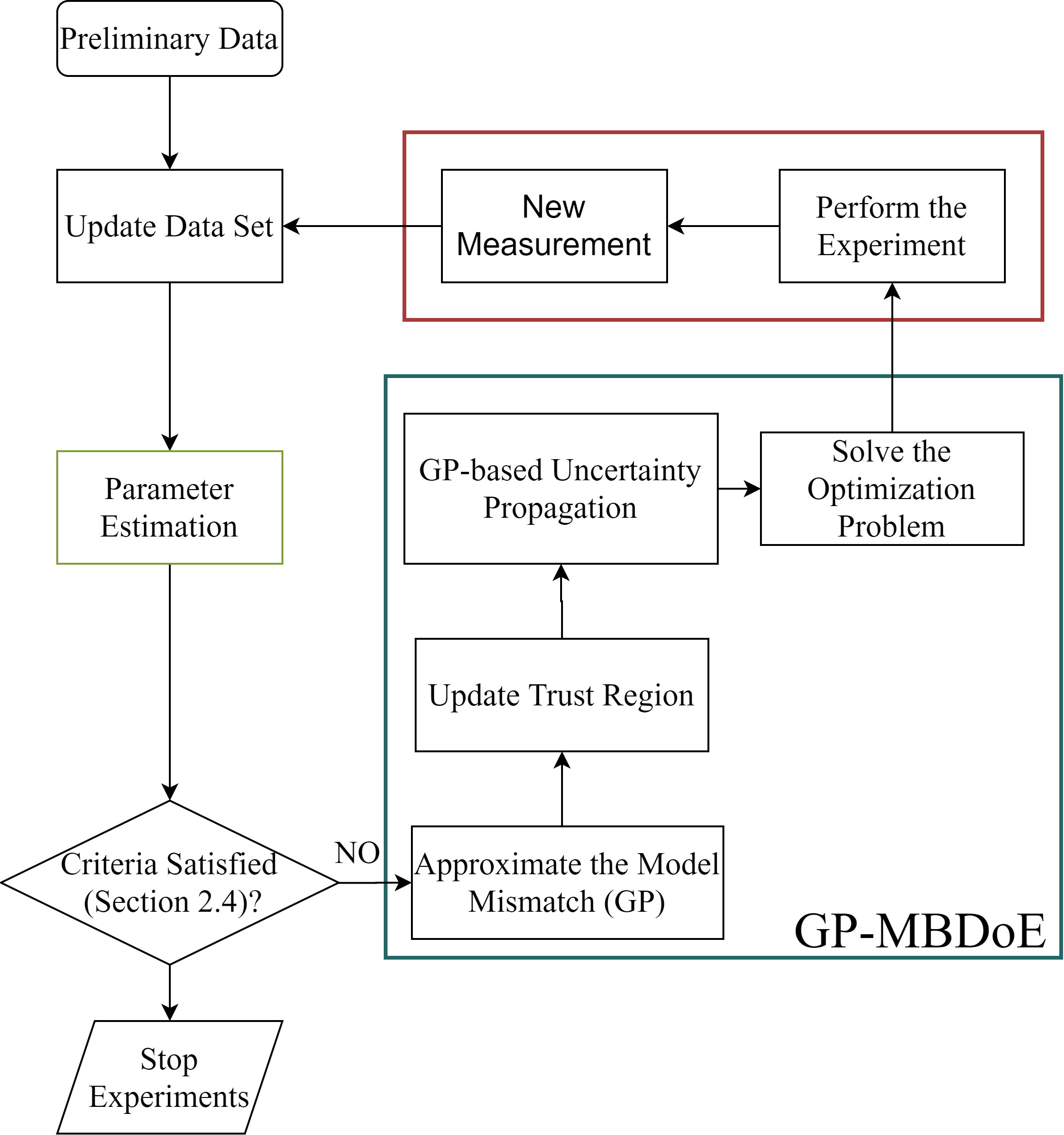}
  \caption{Proposed methodology}
  \label{fig:overall_b}
\end{subfigure}
\caption{Standard (a) and proposed (b) frameworks are presented here, where the standard methodology performs parameter estimation and then with the update kinetic parameters designs the next experiment. On the other hand, theour proposed methodology aims to construct and update trust regions and GP to ensure the chance constraints satisfaction.}
\label{fig:overall}
\end{figure}

A block diagram with the standard framework is given in Fig.~\ref{fig:overall}a. The fundamental steps are briefly illustrated: 
\begin{enumerate}
    \item A preliminary set of data is imported in the system.
    \item Parameter estimation is performed for the approximated model.
    \item The estimated parameters are passed in (cyan) design block, where the new experimental conditions are computed. This step is the core step, where the uncertainty of the parameters together with their corresponding estimates are used to estimate the following experimental conditions.\label{item:block}
    \item The experiments are performed and the data-set is updated
    \end{enumerate}
 This closed-loop is terminated when any of the following holds:
 \begin{itemize}
     \item Statistics are satisfied.
     \item The optimal conditions do not change in between iterations.
 \end{itemize}

\subsection{Termination Criteria.}

The procedures described in Fig. \ref{fig:overall} may terminate when either of the following two conditions are met: i) The design variables do not change within two cycles or ii) the statistics are satisfied. 
The parameters $\hat{\pmb{\theta}}$ and the model predictions that are obtained from the parameter estimation step are associated with the $t$-values and the sum of squared residuals, $\chi^2_{sample}$:
\begin{equation}
    \begin{split}
        &\chi^2_{sample} = \dfrac{(y_{ij}-\hat{y}_{ij})^2}{\sigma_{y_i}^2}\\
        &t_j = \dfrac{\hat{\theta}_j}{\sqrt{\textbf{V}_{\theta_{jj}}}t_{ref}(1-\alpha/2,N_y-N_{\theta})},
    \end{split}
\end{equation}
where $t_{ref}$ is a reference t-value given by a Student t-distribution with $N_y-N_\theta$ degrees of freedom. The model provides an accurate representation of experimental data if $\chi^2_{sample}<\chi^2(1-\alpha, N_y-N_\theta)$, with $\chi^2(1-\alpha, N_y-N_\theta)$ a reference value
from a $\chi^2$ distribution with $N_y-N_\theta$ degrees of freedom and confidence level $1-\alpha$. The model parameter $j$ requires $t_j>t_{ref}(1-\alpha,n_y-n_\theta)$ to be statistically significant.


\section{Methodology}\label{sec:methods}
This section presents the proposed methodology. Gaussian process is used in a two-fold manner: 1) to represent the model mismatch between the available model and the collected data is approximated using GP. To ensure the reliability of the predictions of the GP, trust regions will be applied to constrain the region of experimental conditions where predictions are reliable. 2) An additional GP is then employed to compute the expected value and variance of the objective  (i.e. metric of expected information/variance?) with respect to the parametric uncertainty. 

The intuition behind the two-fold use of GPs is the non-parametric approximation for the physical system to enforce the constraint satisfaction and the same time, maximize the information of the approximated parametric model in expectation. 
Notice that a global GP would not be sufficient since it suffers from the same issues that the approximated confidence regions have; it requires a sufficient number of data-points. If there are enough data-points, the GP can provide a prediction for both the mean and the variance for the approximation. In the presence of limited data points, the GP may result in unreliable predictions unless assumptions are made regarding the function's complexity (see section \ref{sec:methods}). 

\subsection{Proposed framework}

An `ideal' methodology would solve the optimization problem described in (\ref{eqobj_inf}-\ref{eq:constr_inf}). However such an approach is intractable for the following reasons: 1) The real system is not known and 2) the expectation and probabilities are in general intractable integrals.
To accommodate these issues, we propose a novel methodology.
A block diagram with the proposed framework is given in Fig.~\ref{fig:overall}b. The robust MBDoE block in the standard framework (~\ref{fig:overall}a) is replaced by theour proposed method, termed Gaussian process-based MBDoE (GP-MBDoE)

The proposed algorithm is based on GP to approximate the model mismatch between the process and the model, additional trust regions are applied to restrict the feasible design space per iteration. Trust regions are additive constraints for the design variables to limit the distance of the next design from the previous design. After that an additional GP is employed in Fig.~\ref{fig:overall}b block `GP-based Uncertainty Propagation' to approximate the expectation of the metric of the Fisher information and propagate the posterior distribution of the parameters.

The GP-MBDoE performs 4 additional steps before the computation of the new design. 
\begin{itemize}
    \item With the available data, a GP is computed for the plant-model mismatch  (details given in Section~\ref{sec:GP_mismatch}).
    \item A trust region (TR) is updated-computed to restrict the new designs and allows the use of GP locally (details given in Section~\ref{sec:TR}).
    \item The parametric uncertainty (e.g. confidence regions) can be computed via the methods described in section 2.1 in the parameter estimation stage. The GP at this stage aims to approximate the input-output behaviour of the parametric model and not the real system; hence in-silico data points are only employed.  The parametric uncertainty is then propagated to compute the objective function of the design of experiments (more details are given in Section~\ref{sec:Propagation} and \ref{sec:Uncertainty propagation} and in \cite{Olofsson2018}).
    \item The above are then used to perform the optimization step and find the new optimal design with the safety considerations included  (details given in Section~\ref{sec:Optimize}).
\end{itemize}

\subsection{Gaussian Process Fundamentals}
In this section GP is introduced, as they are one of key components of the methodology that follows.
GP generalizes multivariate Gaussian distribution to a distribution over infinite dimensional vector of functions, such that every finite sample
of function values is jointly Gaussian distributed. Due to the Bayesian nature of the solution, GPs are able to consider both epistemic  (limited data) and aleatoric (stochasticity of the true model) uncertainty. A GP is fully specified with the prior mean function $\mu(\cdot)$ and the positive semi-definite kernel function $k(\cdot,\cdot)$, and the GP regression aims to model an unknown set of functions $\mathbf{f}:\mathbb{R}^{n_x}\rightarrow\mathbb{R}^{n_y}$ given some noisy observations ${y}_i=f_i(\textbf{x})+\epsilon_i$, $\epsilon_i\sim \mathcal{N}(0,\sigma_i^2)$. For $\mathbf{f}=\left[f_1,…,f_{n_y}\right]$, this could be expressed as:
\begin{equation}
    f_i\sim \mathcal{G}\mathcal{P}(\mu_i(\cdot),k_i(\cdot,\cdot)),
\end{equation}
where the prior mean $\mu_i$ function provides knowledge of the mean of a test point prior to observing data, the kernel function $k_i$ expresses the covariance between points. Let a number of data points be available, the posterior distribution at a test point $x^*$ is then found by the conditional distribution given on the available $N$ noisy data for each output $i$: $\mathcal{D}_i :=\{\textbf{X},\textbf{y}_i\}$, with $\textbf{X}=\left[\textbf{x}_1,\dots,\textbf{x}_N\right]\in \mathbb{R}^{n_u\times N}$ and $\textbf{y}_i=\left[{y}_{i1},\dots,{y}_{iN}\right]^T\in \mathbb{R}^{N}$. Then, the posterior's mean $m_i({\textbf{x}^*})$  and variance $\Sigma_i({\textbf{x}^*})$ of the test point $\textbf{x}^*$ are
\begin{equation}\label{eq:posterior}
\begin{split}
    &m_i({\textbf{x}^*}) = \mathbb{E}(f_i^*|{\textbf{x}^*},\mathcal{D}) = k_i({\textbf{x}^*},\textbf{X})\left[K_i(\textbf{X},\textbf{X}) + \sigma_i^2 \right]^{-1}(\textbf{y}_i -\mathbf{1}\mu_i({\textbf{x}^*}))+ \mu_i({\textbf{x}^*})\\
    &\Sigma_i({\textbf{x}^*}) = \mathbb{V}(f_i^*|{\textbf{x}^*},\mathcal{D}) = k_i({\textbf{x}^*},{\textbf{x}^*}) - K_i({\textbf{x}^*},\textbf{X})\left[K_i(\textbf{X},\textbf{X}) +\sigma^2\right]^{-1}K_i(\textbf{X},{\textbf{x}^*}),
\end{split}
\end{equation}
Notice that $\Sigma_i$ is the variance of the noise-free prediction, if the noise is taken into consideration then variance should be $\Sigma_i + \sigma_i^2$, and $\sigma_i^2$ is many times treated as a hyperparameter for the GP. The mean and covariance function (kernel) define the prior GP:
\begin{equation}
    \textbf{f}({\textbf{x}^*}) \sim \mathcal{N}(\textbf{m}({\textbf{x}^*}),\pmb{\Sigma}({\textbf{x}^*})),
\end{equation}
with $\textbf{m}({\textbf{x}^*}) =\left[m_1,\dots,m_{n_y}\right]$ and $\pmb{\Sigma}({\textbf{x}^*}) = \text{diag}(\left[\Sigma_1({\textbf{x}^*}),\dots,\Sigma_{n_y}({\textbf{x}^*})\right])$.
Common choices for the kernel $k_i(\cdot,\cdot)$ is the squared-exponential (SE) covariance function:
\begin{equation}
    k_i(\textbf{x},\textbf{x}') =\sigma^{\alpha2}_i \exp\left(-\dfrac{1}{2} (\textbf{x}-\textbf{x}')\pmb{\Lambda}_i (\textbf{x}-\textbf{x}')\right)
\end{equation}
and Mat{\'{e}}rn class of covariance functions, with special cases the  Mat{\'{e}}rn $3/2$ and  Mat{\'{e}}rn $5/2$:
\begin{equation}
\begin{split}
      &k^{3/2}_i(\textbf{x},\textbf{x}') = \sigma^{\alpha2}_i\left(1+\sqrt{3} r_i\right)\exp(-\sqrt{3}r_i)\\
    &k^{5/2}_i(\textbf{x},\textbf{x}') = \sigma^{\alpha2}_i\left(1+\sqrt{5} r_i + \dfrac{5}{3}r_i^2\right)\exp(-\sqrt{5}r_i),
\end{split}
\end{equation}
with $r_i = \sqrt{\left(\textbf{x}-\textbf{x}')\pmb{\Lambda}_i (\textbf{x}-\textbf{x}'\right)}$. The parameters $\sigma^{\alpha2}_i$, $\pmb{\Lambda}_i=\text{diag}([\lambda_1,\dots,\lambda_{n_x}])$ and the noise level $\sigma_i^2$ are hyper-parameters for the GP and, to estimate them, a maximum likelihood estimation is carried out.In this work, we wish to estimate an a priori unknown function $\textbf{f}$ with GP by collecting measurements during the operation. 

\subsection{Gaussian Process for Model Mismatch}\label{sec:GP_mismatch}
In this work, the constraints are not imposed based on the approximated parametric uncertainty, as it is shown that such an approach would suffer in small data sets. For this reason, the model will be used as a prior for the GP that will approximate the model mismatch. Since the real system states affect the feasible space, here we propose approximating constraints instead of the real system as a whole. Specifically, the GP's prior for the constraints is computed using (\ref{eq:approx_inf}) first discretizing it, and then constraints can be computed. As a result, the constraints based on the model are:
\begin{equation}
\begin{split}
    &\hat{\textbf{x}}(t_0) = \textbf{x}(0)\\
   &\hat{\textbf{f}}(\dot{\hat{\textbf{x}}}, \hat{\textbf{x}}, \textbf{u},\pmb{\theta}) = 0\\
    &\hat{g}_i\left(\hat{\textbf{x}},\textbf{u} \right)\in \mathbb{X}, \forall i \in \{1,\dots, n_g\},
\end{split}
\end{equation}
then with some abuse of the notation the symbol $\hat{g}_i$ is used for the constraints associated with the available model:
\begin{equation}
    \hat{g}_i\left(\hat{\textbf{x}},\textbf{u} \right) = \hat{g}_i\left(\textbf{u} \right).
\end{equation}
When, constraints ${g}_i\left(\textbf{u} \right)$ are observed, a GP can be built using as prior the predictions of the model $\hat{g}_i$.
The prior mean $\mu_i$ of the $i-{th}$ constraints is $\mu_i = \hat{g}_i$ and now the hyperparameters of the GP can be found for this hybrid model.

Given there are $N$ available experimental points $\mathcal{D}_i :=\{\textbf{U},\textbf{g}_i\}$, with $\textbf{U}=\left[\textbf{u}_1,\dots,\textbf{u}_N\right]\in \mathbb{R}^{n_u\times N}$ and $\textbf{g}_i=\left[{g}_{i}(\textbf{u}_1),\dots,{g}_{i}(\textbf{u}_N)\right]^T\in \mathbb{R}^{N}$, the mean and variance of the posterior for the (\ref{eq:posterior}) can now be written as:
\begin{equation}\label{eq:posterior_hybrid}
\begin{split}
    &m_{g_i}({\textbf{u}^*}) = \mathbb{E}(g_i^*|{\textbf{u}^*},\mathcal{D}_i) = k_i({\textbf{u}^*},\textbf{U})\left[K_i(\textbf{u},\textbf{U}) + \sigma_i^2 \right]^{-1}(\textbf{g}_i -\mathbf{1}\hat{g}_i({\textbf{u}^*}))+ \hat{g}_i({\textbf{u}^*})\\
    &\Sigma_{g_i}({\textbf{u}^*}) = \mathbb{V}(g_i^*|{\textbf{u}^*},\mathcal{D}_i) = k_i({\textbf{u}^*},{\textbf{u}^*}) - K_i({\textbf{u}^*},\textbf{U})\left[K_i(\textbf{U},\textbf{U}) +\sigma^2\right]^{-1}K_i(\textbf{U},{\textbf{u}^*}),
\end{split}
\end{equation}
Then we wish to satisfy the constraints with a given probability. To satisfy the chance constraints, a distributionally robust reformulation is employed via the Chebyshev-Cantelli  theorem  \cite{Mesbah2014}, (see \ref{thm:Chebyshev}) using the mean and variance of the constraints.
\begin{theorem}\label{thm:Chebyshev}(\cite{Mesbah2014})
Consider a chance probabilistic constraint of the form
\begin{equation}
    \mathbb{P}(q\leq 0)\geq 1-\epsilon, ~~~~\epsilon\in (0,1)
\end{equation}
where $q\in \mathbb{R}^{n_q}$ is some random variable. Let $\mathcal{Q}$ be a family of distributions  with mean $\mu_q$ and variance $\Sigma_q$. Then for any $\epsilon\in(0,1)$, the distributionally robust probabilistic constraint 
\begin{equation}
    \inf_{q\sim \mathcal{Q}}\mathbb{P}(q\leq 0)\geq 1-\epsilon
\end{equation}
is equivalent to the following constraint:
\begin{equation}
    \mu_q + r\sqrt{\Sigma_q}\leq 0,
\end{equation}
with $r=\sqrt{\dfrac{1-\epsilon}{\epsilon}}$.
\end{theorem}
This theorem provides a conservative estimate for the constraints as it holds for the whole family of $\mathcal{Q}$.
\begin{remark}
It should be noted that this theorem was proposed for general constraints. Additionally the parameter $r$ is often used as a design parameter to reduce the conservatism of this formulation, however loosing any respected guarantees.  
\end{remark}
This reformulation relies highly on the accuracy of the GP's mean and variance and the initial small amount of data points may significantly affect the result. Notice that even though the result in Theorem \ref{thm:Chebyshev} is conservative for all the family of distribution with mean $\mu_q$ and $\Sigma_q$, the wrong estimation of mean and variance may lead to infeasible designs.
\subsection{Trust Regions}\label{sec:TR}

In this work, different GPs are utilized to tackle the objective function's parametric uncertainty and the structural mismatch between the model and process. This section focuses on how the GP can be trained and provide guarantees for the latter case. Mainly, trust regions are utilized to update the GPs for the model-mismatch locally and not globally.
First, we introduce the assumption and the bound that it is usually employed, then we show that these bounds may not be used globally using a motivating simple example.

It is common to achieve bounds for the unknown function, using the following assumption:
\begin{assumption} \label{Assume_fun}
\cite{Sui2015, Berkenkamp2017} The unknown function ${f}_i$ has bounded norm in reproducing kernel Hilbert space (RKHS) $\mathcal{H}_k$ \cite{Scholkopf}, induced by a the continuously differentiable kernel $k_i$, $||{f}_i||_{k_i}\leq B_{f_i}$, $B_{f_i}>0$.
\end{assumption}
Note that Assumption~\ref{Assume_fun} does not make any probabilistic assumption on $f_i$ but it requires to have a bounded norm in the RKHS or be drawn by a GP prior \cite{Srinivas2010}. 
Given assumption~\ref{Assume_fun}, reliable confidence intervals on $f_i$ can be found:

\begin{lemma}\label{conf_bound}\cite{Srinivas2010}
Let a function $f_i$ that satisfies Assumption  \ref{Assume_fun} with $k_i$, $||{f}_i||_{k_i}\leq B_{f_i}$ and the measurement noise is a $\sigma-$sub-Gaussian noise. Then  $|m_i(\textbf{x})-g_i(\textbf{x})|\leq \beta^{(N)}_i\sigma_i(\textbf{x})$ with probability higher than $1-\delta$, and  $\beta_i^{(N)} = B_{f_i} +4 \sigma \sqrt{\gamma_N+1-\log(\delta)}$, with $\gamma_N$ being the information capacity of $k_i$.
\end{lemma}

The scaling factor $\beta_i^{(N)}$
in Lemma \ref{conf_bound} depends on the kernel's $k_i$ information capacity $\gamma_N$. It is the maximum amount of mutual information \cite{Srinivas2010} that can be achieved using the noisy measurement $\mathcal{D}_i$ for $f_i$. 
Due to generic nature of the systems, the Assumption~\ref{Assume_fun} may not hold globally. For that reason we will aim to use the GP locally instead. 

To motivate the need of local approximations and trust regions the following example is presented.

\subsubsection{Importance of Trust Regions: A simple motivating example }
We first provide a simple motivating example that illustrates the reasons why such approximation is not cautious globally, and then the trust regions for our problem are introduced.

\begin{figure}[H]
    \centering
    \includegraphics[scale=0.5]{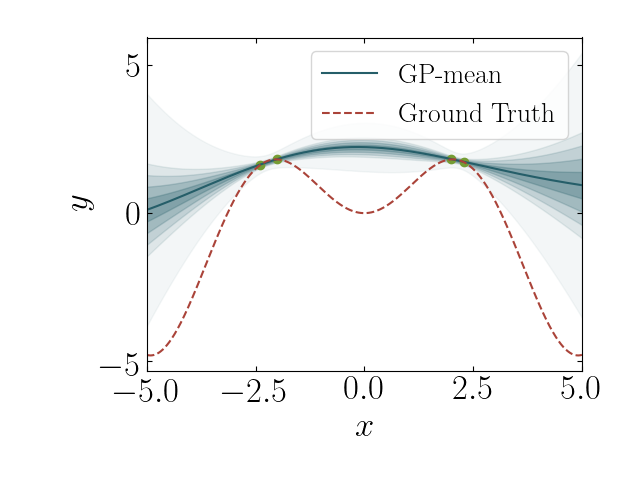}
    \caption{The ground truth (red dashed line) is the function that the GP approximates given the available data. The GP predictions for the mean ($\mu$) is the blue line with variance $\Sigma$, and the shaded areas represents the area between $m\pm\sqrt{\Sigma}, m\pm\sqrt{\Sigma}, \mu\pm3\sqrt{\Sigma}, m\pm4\sqrt{\Sigma}$ and  $m\pm10\sqrt{\Sigma}$.}
    \label{fig:simple_GP_danger}
\end{figure}

Let a simple function $f(x) = x \sin(x)$ be assumed as the `true' model, with $x\in \mathbb{R}$ and the collected data are corrupted by a Gaussian noise $\epsilon_f$ with standard deviation 0.01, $y=f(x) + \epsilon_f$. A GP is used to approximate the data points, specifically a squared-exponential (SE) covariance function is used and its hyperparameters are fitted using maximum likelihood estimation (with multi-starts  to avoid local optima). The mean $m$ and variance $\Sigma$ of the GP for each $x$ are illustrated in Fig~\ref{fig:simple_GP_danger}. 

The shaded areas show the $m+r \sqrt{\Sigma}$ for $r=1, 2, 3, 4$ and $10$. Using the Theorem~\ref{thm:Chebyshev} and the Lemma~\ref{conf_bound}, we would expect the  `true' model (red dashed line) to lie within these shaded areas. Specifically, the Chebyshev-Cantelli theorem states that for $r=9.94$,  $x$ should lie within the shaded area with probability $99\%$. However, due to the shape of the  `true' model and the collected data, such guarantee is lost as the predicted variance and mean are poorly computed. This motivates the further restriction of the proposed optimization problem, as it is clear that the GP cannot be used globally with small data-sets and it is a paramount importance to satisfy the constraints with high probability.
%
%
%
\subsubsection{Trust region updates}
The use of local models has extensively been employed in the literature of derivative-free optimization~\cite{Conn2005, Cartis2011}, where trust regions (TRs) are used to bound the predictions of the surrogate function of the respected objective function around the current point. The size of the TR is updated in every iteration according to a ratio of the objective's actual reduction, over the predicted reduction. In this work the same philosophy is used for the approximation of the constraints: A TR is defined for each constraint in the problem and it is updated according to the accuracy of the predictions on the new observations. The TR is defined as:
\begin{equation}\label{TR}
    (\textbf{u}-\textbf{u}_k)^T(\textbf{u}-\textbf{u}_k)\leq \mathcal{R}_i, ~~~\forall i\in\{1,\dots,n_g\}
\end{equation}
where $\textbf{u}_k$ is the operating point (experimental condition) and $\mathcal{R}_i$ the radius of the trust region.
The rule for updating the TR is described in Algorithm \ref{alg:TR}.
\begin{algorithm}[H]
\caption{Trust region updates}\label{alg:TR}
\smallskip
{\bf Input:} Give parameter $\eta_1$, $\eta_2$, $t_2<1<t_1$ and $R_i$. New observation $g_i(\textbf{u}^*)$ and model prediction $\hat{g}_i(\textbf{u}^*)$ for $\textbf{u}^*$.
\smallskip
\begin{enumerate}
\item Compute $\rho_i = \left(\dfrac{(g_i(\textbf{u}^*)-\hat{g}_i(\textbf{u}^*)}{g_i(\textbf{u}^*)}\right)^2$
\item If $\rho_i\leq \eta_1$: $\mathcal{R}_i = t_1\mathcal{R}_i$
\item If $\rho_i\geq \eta_2$: $\mathcal{R}_i = t_2\mathcal{R}_i$
\item If $\eta_1\leq\rho_i\leq \eta_2$: $\mathcal{R}_i = \mathcal{R}_i$
\end{enumerate}
{\bf Output:} New radius $\mathcal{R}_i$.
\end{algorithm}
According to Algorithm~\ref{alg:TR}, the radius $\mathcal{R}_i$ is updated according to the accuracy ($\rho_i$) of the new predicted point. If the percentage error is high, i.e. $\rho_i\leq \eta_1$ then the trust region is reduced if the  percentage error is low then the trust region is increased.

The values of the parameters $\eta_1$, $\eta_2$, $t_2<1<t_1$, can be seen as notion of conservatism for the optimization. If the parameters $t_1$ and $t_2$ are closer to $1$, then the scheme becomes more conservative, as the trust region changes very little at each iteration. The values for  $\eta_1$ and $\eta_2$ are the bounds that we want for the approximation of the physical system. If their values are selected to be large then relaxation is imposed to the scheme, as the radius will increase even if the GP is not good enough. 

The predictions are now inside the trust region and the computed mean and variance are accurate locally. This way the Lemma~\ref{conf_bound} and Theorem~\ref{thm:Chebyshev} may be applied to guarantee the feasibility of the experimental design.

\subsection{GP-based Uncertainty Propagation for the Information Metric}\label{sec:Propagation}
The GPs are further utilized to compute the metric of information/variance accounting for the parametric uncertainty. 
GP will be used to directly approximate the metric as function of the optimization variables and the model parameters. 
Then, model parameter marginalization is used to compute the expected value and variance of the information metric with respect to a computed uncertainty of the parameters.
The use of surrogates to propagate the uncertainty is not new in MBDoE, where polynomial chaos expansion is mainly used.
Similar approach was followed in \cite{Olofsson2018}, but for the case of model discrimination. 

Consider the objective function of the experimental design at the $k^{th}$ experiment: 
\begin{equation}
\begin{split}
   J(\textbf{u},\pmb{\theta}) =& \psi\left(\dfrac{\partial \hat{\textbf{x}}(t_N)}{\partial \pmb{\theta}}^\top\Sigma_{exp}\dfrac{\partial \hat{\textbf{x}}(t_N)}{\partial \pmb{\theta}}+\textbf{M}_0\right)\\
  & s.t.\\
  &\hat{\textbf{x}}(t_0) = \textbf{x}(0)\\
   &\hat{\textbf{f}}(\dot{\hat{\textbf{x}}}, \hat{\textbf{x}}, \textbf{u},\pmb{\theta}) = 0\\
        &\hat{\textbf{y}} = \hat{\textbf{h}}(\hat{\textbf{x}},\textbf{u})
\end{split}
\end{equation}

Notice that in this case the training set for the GP is generated in-silico using the available approximated model, hence a sufficiently large data-set can be generated \cite{Olofsson2018}. This is a main difference in the training of this GP with the ones used for the model mismatch, as there is no need for local approximations.
Given the uncertainty region that has been computed (see section \ref{sec:mle}) the uncertainty can not be directly propagated via the GP.
The posterior distribution will not be Gaussian and a Monte-Carlo sampling is needed.
For that reason further approximations are utilized; a very common approximation is to use Taylor expansion (see ~\ref{sec:Gaussian_approxim} with an additional example), where the expected value and variance of the objective is computed with respect to the uncertain parameters $\pmb{\theta}$.

The expected value of the objective in MBDoE (the information metric) can be substituted by the mean function of the GP. Note that the objective function can also incorporate a penalty term for large variations \cite{Bradford2018b}. The objective function is now reformulated as:
\begin{equation}\label{eq:obj_propagate}
    \hat{m}_J(\textbf{u},\pmb{\mu}_{\hat{\pmb{\theta}}},\pmb{\Sigma}_{{\hat{\pmb{\theta}}}}) + \alpha_J \sqrt{\hat{\Sigma}_J(\textbf{u},\pmb{\mu}_{\hat{\pmb{\theta}}},\pmb{\Sigma}_{{\hat{\pmb{\theta}}}})},
\end{equation}
where $\hat{m}_J(\textbf{u},\pmb{\mu}_{\hat{\pmb{\theta}}},\pmb{\Sigma}_{{\hat{\pmb{\theta}}}})$ and ${\hat{\Sigma}_J}(\textbf{u},\pmb{\mu}_{\hat{\pmb{\theta}}},\pmb{\Sigma}_{{\hat{\pmb{\theta}}}})$ are the mean and variance of the GP after the propagation of the uncertainty of the parameters $\pmb{\theta}$, $\alpha_J$ is a constant variable. The value of  $\hat{m}_J(\textbf{u},\pmb{\mu}_{\hat{\pmb{\theta}}},\pmb{\Sigma}_{{\hat{\pmb{\theta}}}})$ represents the information metric in expectation when uncertain parameters have mean $\pmb{\mu}_{\hat{\pmb{\theta}}}$ and $\pmb{\Sigma}_{{\hat{\pmb{\theta}}}}$.

\subsection{Approximate Optimal Model-Based Design of Experiments}\label{sec:Optimize}
The optimization problem that is solved to fnd the new design $\textbf{u}^*$ given the previous design $\textbf{u}_k$ and the parameter's computed mean ($\pmb{\mu}_{\hat{\pmb{\theta}}}$) and variance ($\pmb{\Sigma}_{{\hat{\pmb{\theta}}}}$) is:
\begin{equation}\label{eq:opt}
    \begin{split}
        \min_\textbf{u}~~& \hat{m}_J(\textbf{u},\pmb{\mu}_{\hat{\pmb{\theta}}},\pmb{\Sigma}_{{\hat{\pmb{\theta}}}}) - \alpha_J \sqrt{\hat{\Sigma}_J(\textbf{u},\pmb{\mu}_{\hat{\pmb{\theta}}},\pmb{\Sigma}_{{\hat{\pmb{\theta}}}})}\\
        &s.t.\\
        & {m}_{g_i}(\textbf{u}) + r\sqrt{\Sigma_{g_i}(\textbf{u})}\leq0\\
        &(\textbf{u}-\textbf{u}_k)^T(\textbf{u}-\textbf{u}_k)\leq \min(\mathcal{R}_i)\\
        &\forall i\in\{1,\dots,n_g\},
    \end{split}
\end{equation}
where $3>\alpha_J>0$ is used to add exploration to the design. This parameter can be seen as the upper confidence bound formulation in Bayesian optimization \cite{frazier2018tutorial}. The larger the value of $\alpha_J$ the more significant the exploration will be. In this work, the optimization will mainly focus on exploitation, and for that reason $\alpha_J$ will have the value of $0.1$. If the user wants to emphasize to exploration then values between 2 and 3 should be used, inspired by the upper confidence bound methods \cite{Jones1998}.

The Algorithm \ref{alg:approximateOED} summarizes the proposed framework.
\begin{algorithm}[H]
\caption{Approximate optimal model-based design of experiments}\label{alg:approximateOED}
\smallskip
{\bf Input:} Give parameter $\eta_1$, $\eta_2$, $t_2<1<t_1$, $\alpha_J$ and $tol_1$. For $i^{th}$ constraint give $N$ data  $\mathcal{D}_i :=\{\textbf{U},\textbf{g}_i\}$, with $\textbf{U}=\left[\textbf{u}_1,\dots,\textbf{u}_N\right]\in \mathbb{R}^{n_u\times N}$ and $\textbf{g}_i=\left[{g}_{i}(\textbf{u}_1),\dots,{g}_{i}(\textbf{u}_N)\right]^T\in \mathbb{R}^{N}$ and radius $R_i$. Posterior distribution for the model parameters $\Theta$
\smallskip
\begin{enumerate}
\item Compute the model predictions for each constraint, $\hat{g}_i$ for all $\textbf{U}$.
\item Train GP for the constraints using $\hat{g}_i(\textbf{u})$ as the prior.
\item Generate in-silico data for the objective function: $J:\textbf{U}\times{\Theta}\rightarrow \mathbb{R}$

\item Train GP to approximate $J$ and propagate uncertainty using equations (\ref{eq:gaussian_aprx}-\ref{eq:gaussian_aprx_forms}) \& (\ref{eq:obj_propagate})
\item Solve Optimization~(\ref{eq:opt}) that results in $\textbf{u}^*$
\end{enumerate}
{\bf Output:} Get new design $\textbf{u}_{k+1} = \textbf{u}^*$.
\end{algorithm}

The steps that followed to solve the MBDoE problem using GPs are given below: First in \textbf{step 1} and \textbf{step 2} a Gaussian process is trained to approximate the constraints for the physical system. Here experimental data gathered from the process are used. Notice that first the model predictions for the model are computed as the model is used as prior for the GP (see Section \ref{sec:GP_mismatch}). 
Then, in-silico artificial data are generated in \textbf{step 3} to approximate the objective function using a GP in \textbf{step 4} (see Section \ref{sec:Propagation} and \ref{sec:Gaussian_approxim}). In \textbf{step 4}, the uncertainty is propagated (see Section \ref{sec:Propagation} and \ref{sec:Gaussian_approxim}). Finally the optimization is solved in \textbf{step 5}, using  (\ref{eq:opt}). In this step, the trust regions are applied, in order to restrict the feasible space and maintain a good local approximation.

The proposed framework, GP-MBDoE, first shown in Fig~\ref{fig:overall}b can now be described in Algorithm~\ref{alg:overall}. Algorithm \ref{alg:overall} represents the proposed framework, whereas the Algorithm \ref{alg:approximateOED} is utilized at the step of that the next optimal experimental design.

\begin{algorithm}[H]
\caption{Proposed Framework}\label{alg:overall}
\smallskip
{\bf Input:} Give parameter $\eta_1$, $\eta_2$, $t_2<1<t_1$, $\alpha_J$ and $t_3<1$. For $i^{th}$ constraint give $N$ data  $\mathcal{D}_i :=\{\textbf{X},\textbf{y}_i\}$, with $\textbf{U}=\left[\textbf{u}_1,\dots,\textbf{u}_N\right]\in \mathbb{R}^{n_u\times N}$ and measurements $\textbf{g}_i=\left[{g}_{i}(\textbf{u}_1),\dots,{g}_{i}(\textbf{u}_N)\right]^T\in \mathbb{R}^{N}$ for each constrained $i$,
and radius $\mathcal{R}_i$.
\smallskip

Select design $\textbf{u}_0$\\
\textbf{for} k = \{1,\dots\}:
\begin{enumerate}
\item Perform Parameter estimation to obtain $\hat{\pmb{\theta}}$.
\item Approximate prosterior distribution $\Theta$ and compute mean $\pmb{\mu}_{\hat{\pmb{\theta}}}$  and variance $\pmb{\Sigma}_{{\hat{\pmb{\theta}}}}$.
\item  \textbf{if} $||\textbf{u}_{k+1}-\textbf{u}_k||\leq tol_1$ \textbf{or} statistics are satisfied:\\
~~~~~~~~~Exit
\item Find the new design $\textbf{u}_{k+1}$ using  \textbf{Algorithm~\ref{alg:approximateOED}} with inputs($\textbf{U}, \textbf{g}_i,\pmb{\mu}_{\hat{\pmb{\theta}}},\pmb{\Sigma}_{{\hat{\pmb{\theta}}}},\mathcal{R}_i,\alpha_J$).
\begin{enumerate}[label=(\roman*)]
    \item Update $\mathcal{R}_i$ using \textbf{Algorithm {\ref{alg:TR}}} with inputs ($t_1,t_2,\eta_1,\eta_2,\mathcal{R}_i$).
    \item \textbf{if} $g_i(\textbf{u}_k)>0$: $\mathcal{R}_i=t_3\mathcal{R}_i$ and $\textbf{u}_{k+1} = \textbf{u}_k$
\end{enumerate}
\item Perform new experiment and update $\textbf{U}, \textbf{g}_i$ and $\textbf{y}_j$.

\end{enumerate}
{\bf Output:} Experiments finished.
\end{algorithm}

The steps followed in this work as illustrated in Fig.~\ref{fig:overall_b} are the following for each iteration $k$: Given the preliminary data (and the updated data), \textbf{Parameter Estimation} is performed, specifically
$\textbf{Step 1}$ and $\textbf{Step 2}$ are used to obtain the best parameters $\pmb{\theta}$ given the initial data points we have and also approximate the posterior distribution. As it is described in section \ref{sec:mle}, the parametric uncertainty that is obtained is not suitable for the constraint satisfaction. However, the parametric uncertainty can be used for the objective of the experimental design.  If the \textbf{Termination criteria} are satisfied (see Section 2.4)  then in \textbf{Step 3} the experimental procedure stops. Algorithm 2 is then performed in \textbf{Step 4}, where the new optimal design is found. This step also entails the construction of the GP for the model mismatch  using the parametric model as prior  (\textbf{Approximate the Model Mismatch} in Fig.~\ref{fig:overall_b}), the GP-based uncertainty propagation for the objective function (\textbf{GP-based Uncertainty Propagation} in Fig.~\ref{fig:overall_b}) and solution of the optimization for the computation of the new design (\ref{eq:opt}) (\textbf{Solve the Optimization Problem} in Fig.~\ref{fig:overall_b}). 
In \textbf{Step 4.i-ii} the \textbf{Update Trust region} (Fig.~\ref{fig:overall_b}) step is performed following Algorithm~\ref{alg:TR}. After that in \textbf{Step 5} the new experiment is performed and the last step contains a backtracking that is used when an infeasible design happen where the next design is return to the previous value and the trust region of that constraint shrinks. 

\section{Case Studies}
The proposed methodology (GP-MBDoE) is illustrated using two case studies with different scenarios simulated in silico: 
\begin{itemize}
    \item Case Study 1: a plug flow reactor where that the assumed mechanism is different than the `real' one with an additional constant disturbance. The model contains 4 parameters and has 2 decision variables.
    \item Case Study 2: a plug flow reactor that a Nucleophilic aromatic (SnAr) substitution \cite{Hone2017} reaction takes place, where the concentrations have a non-constant disturbance affect by the inlet concentrations. Such phenomenon can be observed in case a pump malfunction occurs. The model contains 8 parameters and has 4 decision variables.
\end{itemize}
For each case study, two models structures are postulated: (i) a structure to simulate the physical system, which is employed to perform the in silico experiments and (ii) a misspecified model structure with corrupted measurements. Once a preliminary instance of the model parameters for each of the structurally incorrect models is obtained, both case studies involve an experimental design stage. In order to evaluate our methodology, we compare the GP-MBDoE with the standard methods that consider uncertainty: 
\begin{itemize}
    \item Monte Carlo-based robust MBDoE (MC-MBDoE) under parametric uncertainty, where the Monte Carlo using sampling from the posterior distribution of the parameters to calculate the backoffs \cite{Galvanin_backoff}. A sampling of the expected uncertainty domain of the parameters needs to be performed to compute the effect of the possible realizations of the unknown parametric set on the model's state variables. Stochastic simulations are performed to evaluate the uncertain parameter's effect and compute the tightening of the constraints (backoffs). The optimization is solved, and the percentage of the constraint violation is computed, and the backoffs are re-estimated, and the stochastic simulation is performed. This process continues until the backoffs have converged. 
    Notice that other more sample efficient methods could be used \cite{Bradford2018b}, but would require additional assumption for the distribution of the constraints.
    \item Disturbance estimation-based robust MBDoE (DE-MBDoE), where the structural model mismatch will be modeled by a disturbance, a method that is originated by the MPC literature \cite{MPC_dist} and applied to online MBDoE \cite{Galvanin2012_DIST}. This method can be seen as a simplification to our approach where instead of using a GP, a constant disturbance is used and updated at each step. Estimation of a constant disturbance is performed using the difference between the measurement and predictions. 
\end{itemize}
The GP-MBDoE required to construct different GPs for the objective function and the constraints. The GP for objective function requires simulations of the approximated model.
For that reason, a sufficiently high number of data points should be used.
The kernel used for the objective function is the Mat{\'{e}}rn $5/2$ class of covariance function with 200 and 400 points for each case study, respectively.
For the computational times see (Section \ref{sec:Discussion}).
For the approximation of the model mismatch, where only a limited number of data points can be used, the simple SE covariance function is implemented. 
\subsection{Case Study 1: Flow Reactor: `Wrong Mechanism'}
A flow reactor is assumed where the system can be described by the following set of DAEs:
\begin{equation}
    \dfrac{-F}{A}\dfrac{dc_i}{dz} + \sum_{j=1}^{N_r}a_{ij}r_j =0,
\end{equation}
where $c_i$ is the concentration (mol/L) of the $i^{th}$ reagent, $a_{ij}$ is the stoichiometric coefficient of the $i^{th}$ reagent in the $j^{th}$ reaction with $r_j$ reaction rate. The length of the reactor is $L = 25 \text{cm}$ with surface $A = 1.2 \text{cm}^2$ and  $F (\text{mL}/\text{min})$ is
the inlet flowrate $F$. The temperature dependency of the reaction rate is described using Arrhenius law:
\begin{equation}\label{eq:Arr}
    k_j^r =A_j \exp\left(\dfrac{-E_{\text{a}j}}{R} \dfrac{1}{T+273.15}\right),
\end{equation}
where $A_j$ denotes the frequency factor and $E_{\text{a}j}$ the activation energy. In this setting the design variables are the inlet flowrate $F$ and temperature $T$, i.e. $\textbf{u}= \left[F, T\right]$. The formulation in (\ref{eq:Arr}) tends to be sloppy \cite{BUZZIFERRARIS20091061}, for that reason the following formulation used instead:
\begin{equation}\label{eq:Arr2}
    k_j^r =\exp\left(k_j^0 -\dfrac{{E}_j10^4}{R} \left(\dfrac{1}{T+273.15}- \dfrac{1}{90+273.15}\right)\right),
\end{equation}
The mechanism that takes place in the true model is
$\left\{
\begin{matrix*}[l]
\ce{A} \xrightarrow{k_1} \ce{B} \xrightarrow{k_2} \ce{C}\\
\ce{A} \xrightarrow{k_3} \ce{C}\end{matrix*}
\right\}$, there is an additional constant disturbance ${w}$ of 0.1 M  added to the concentrations. The model that is assumed in the design has the following mechanism $\left\{\ce{A}\xrightarrow{k_1} \ce{B} \xrightarrow{k_2} \ce{C}\right\}$, where one reaction is ignored ($\left\{\ce{A} \xrightarrow{k_3} \ce{C}\right\}$). We assume to have upper and lower bounds on concentration measurements  of $c_{\ce{C}}$, i.e. $g_1=c_{\ce{C}} - 0.4 \leq 0~\text{mol/L}$ and $g_2=c_{\ce{C}} - 0.1 \geq 0~\text{mol/L}$ to be satisfied, satisfied with probability $90\%$, respectively. The measurements are corrupted by normally distributed noise with standard deviation $\left[ \sigma_{\ce{A}},\sigma_{\ce{B}},\sigma_{\ce{C}} \right]=\left[0.039, 0.14, 0.05\right]$. Additionally, reaction rate $i$ has the following form: 
\begin{equation}
    k_j^r = k^0_j \exp\left(\dfrac{-E_{\text{a}j}}{R} \dfrac{1}{T+273.15}\right)
\end{equation}

The parameters for the true model are depicted in Table \ref{tab:real}:

\begin{table}[H]
  \centering
  \caption{Parameter values assumed for the true kinetic model}
    \begin{tabular}{cc}
  \hline			
    $Parameter$ & Value \\
    \hline			
    $k_1^0 (\text{min}^{-1})$  & 8 \\
    $E_{\text{a}1} (\text{kJ/mol})$  & 29 \\
    $k_2^0 (\text{min}^{-1})$  & 5 \\
    $E_{\text{a}2} (\text{kJ/mol})$   & 35 \\
    $k_3^0 (\text{min}^{-1})$  & 3 \\
    $E_{\text{a}3} (\text{kJ/mol})$  & 32\\
  \hline			
    \end{tabular}%
  \label{tab:real}%
\end{table}%
The concentration of the the inlet reagent $\ce{A}$ is constant at $2~\text{mol/L}$. The experimental design variables $T, F$ are bounded with $T(^oC) \in \left[60,100\right]$ and $F(\text{mL/min})  \in \left[0.004,0.008\right]$. Latin hypercube is used to generate an initial set of 5 experiments (Table \ref{tab:initial experiments2}).

\begin{table}[H]
  \centering
  \caption{Initial set of experimental conditions for case study 1}
    \begin{tabular}{cc}
  \hline			
    $T (^o\text{C})$ & $F (\text{mL}/\text{min})$ \\
    \hline			
    79.6  & 0.0069 \\
    97.6  & 0.0055 \\
    62.8  & 0.0077 \\
    88.0  & 0.0048 \\
    69.0  & 0.0062 \\
  \hline			
    \end{tabular}%
  \label{tab:initial experiments2}%
\end{table}%

The following configurations are discussed:\\
\begin{itemize}
    \item  MBDoE under parametric uncertainty (MC-MBDoE).
    \item  MBDoE with disturbance estimation (DE-MBDoE)
    \item  GP-based MBDoE (GP-MBDoE)
\end{itemize}

\begin{figure}[H]
    \centering
    \includegraphics[scale=0.5]{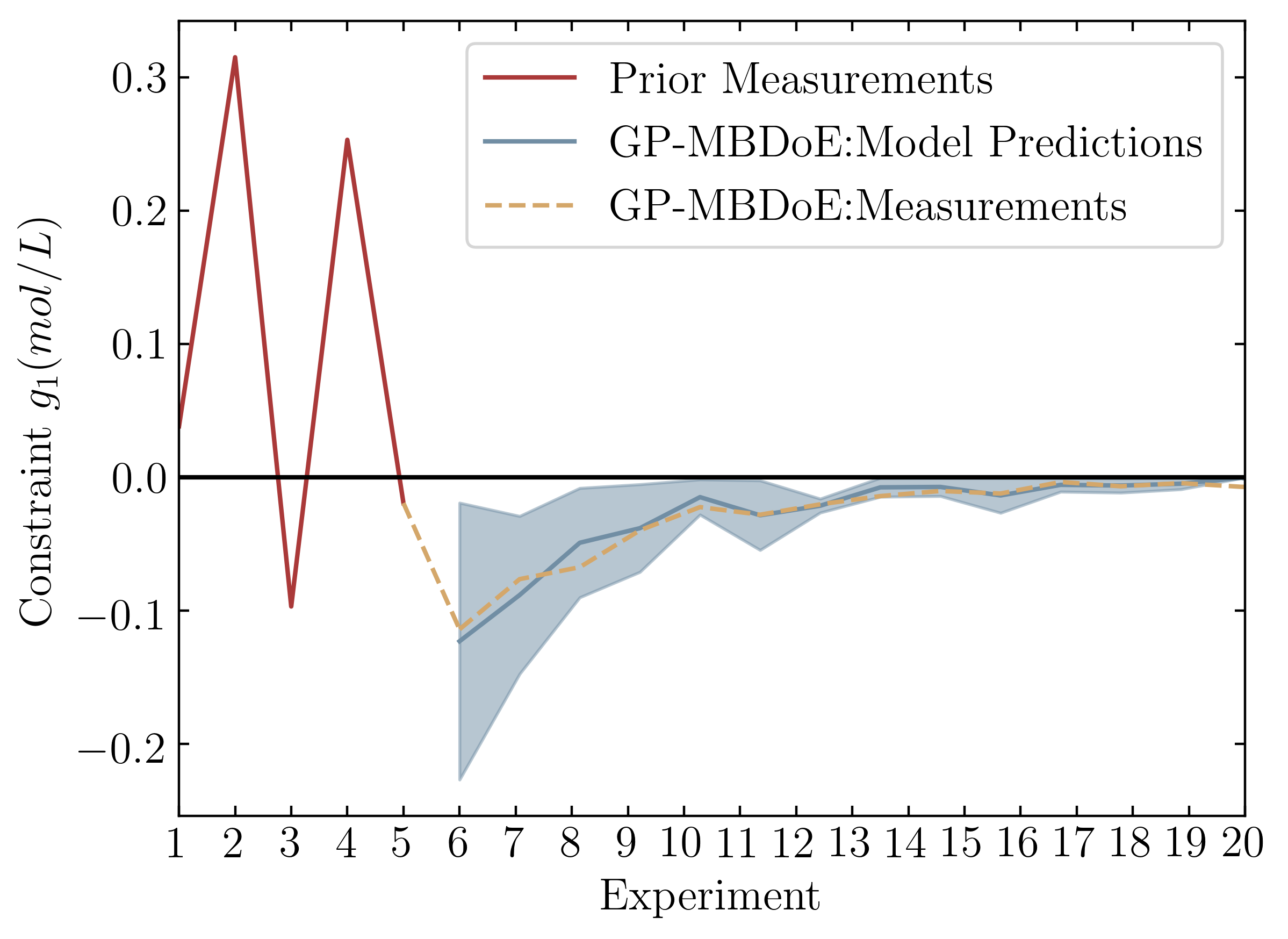}
    \caption{Case study 1: Performance of GP-MBDoE. Blue shaded area represents the 99.7\% confidence interval of the GP. The dashed yellow line is the behaviour of the true model and the red lines represent the value for the constraint at the initial experiments.}
    \label{fig:GP-MBDoE-constraint}
\end{figure}

\begin{figure}
    \centering
    \includegraphics[scale=0.4]{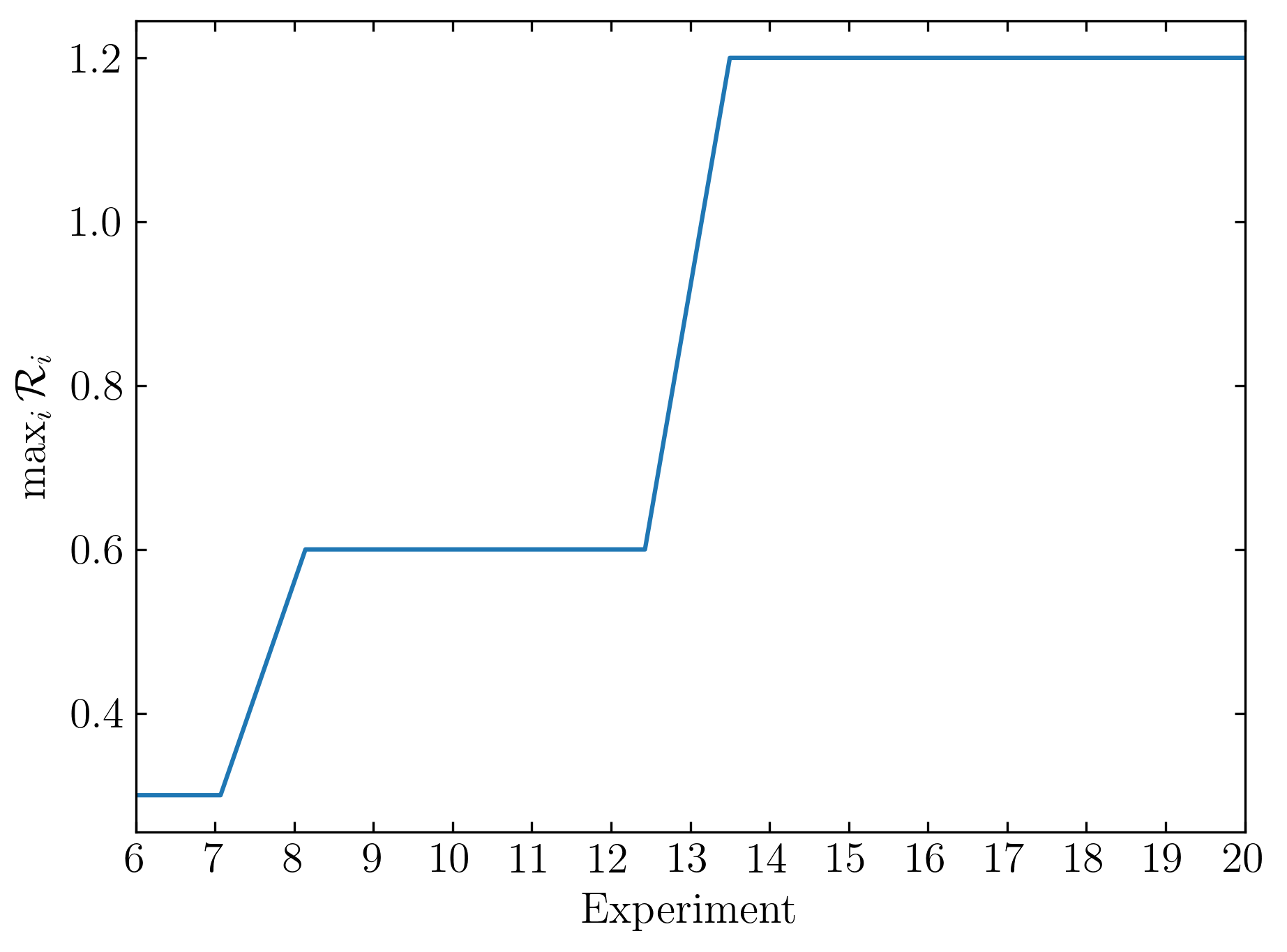}
    \caption{Trust region for the constraint $g_1$}
    \label{fig:TR1_simple}
\end{figure}

A D-optimal design criterion is used in all the proposed experiment design configurations. The results follow in the next paragraphs of this section regarding manipulated inputs, simulated profiles and a posteriori statistics on the final parameter estimation.
To fairly evaluate all the methodologies,  the same method for propagating uncertainty in the objective function is used. Specifically, the posterior distribution of the parameters is approximated with a normal distribution, updated in every iteration.
The parameters needed for the Algorithm \ref{alg:overall} are depicted in Table \ref{tab:opt1}: 
\begin{table}[H]
  \centering
  \caption{Parameters for Algorithm \ref{alg:overall}}
    \begin{tabular}{cc}
  \hline			
    $Parameter$ & Value \\
    \hline			
    $\alpha_J$  & 0.5 \\
    $\eta_1$  & $10^{-3}$ \\
    $\eta_2$  & $10^{-2}$ \\
    $t_2$   & 0.5 \\
    $t_1$  & 2 \\
    $t_3$  & 0.5\\
    $\mathcal{R}_i$ & 0.3\\
  \hline			
    \end{tabular}%
  \label{tab:opt1}%
\end{table}%
Typical values for $t_i$ and $\mathcal{R}_i$ can be found in \cite{Conn2005}. Notice that the trust region is defined for the normalized design variables that are inside $[-1,1]$.
All the methods are applied in the same closed loop manner, and the results are presented next.  Fig. \ref{fig:GP-MBDoE-constraint} depicts the constraint satisfaction of the first constraint $g_1$ for the GP-MBDoE, it is evident that the constraint remain feasible in the all the sequence of experiments even though it approaches the bound of the constraint. The blue shaded area represents the 99.7\% confidence intervals of the GP. As expected, due to the design space that is used in every consecutive optimization, the measurements of the true model lie within the confidence interval. Notice that the confidence interval shrinks as more measurements are acquired. 
This behaviour can also be seen in Fig.~\ref{fig:TR1_simple} , where the radius of trust region $\mathcal{R}_1$ increases as new measurements are acquired and the GP prediction improves. 

\begin{figure}[H]
    \centering
    \includegraphics[scale=0.2]{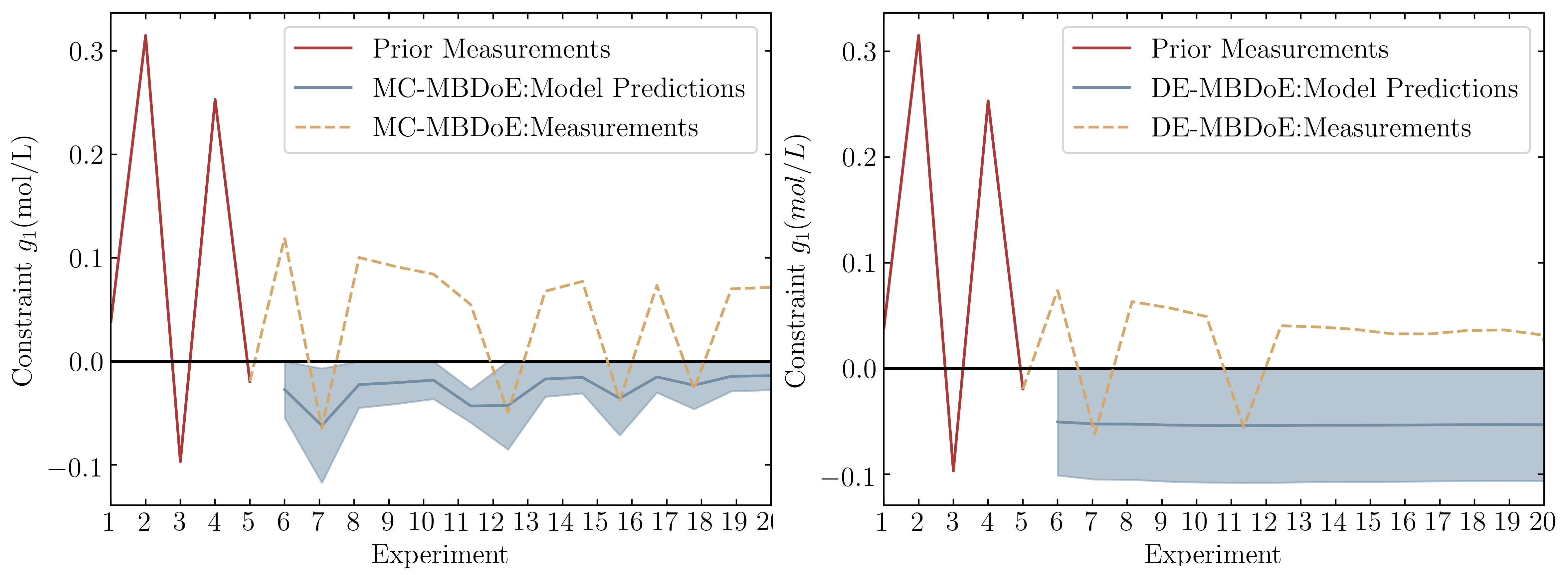}
    \caption{Case study 1: Performance of MC-MBDoE and DE-MBDoE. Blue shaded area represents the computed backoff for each case. The dashed yellow line is the behaviour of the true model and the red lines represent the value for the constraint at the initial experiments.}
    \label{fig:rest-constraint}
\end{figure}

On the other hand none of MC-MBDoE and DE-MBDoE formulations managed achieve the desired goal; i.e. remain in the feasible area of the constraints. This is not a surprise as none of the aforementioned method accounted the structural model mismatch properly. It is noticeable from Fig. \ref{fig:rest-constraint} that the resulted solution from the optimization is predicted to remain feasible with relatively large confidence intervals. However, the MC-MBDoE can account only for a parametric uncertainty, in which the posterior of the parameters has been approximated, and the DE-MBDoE allows the structural model mismatch to be constant that is updated at each iteration. Additionally, in this work Monte-Carlo method is used to solve the stochastic MBDoE problem, which is computationally intensive as in \cite{Galvanin_backoff}; however approximated solutions can be utilized, e.g. polynomial chaos expansions \cite{Mesbah2014}. 

Notice that the design with GP-MBDoE defines a test that is now both safe (feasible) and informative for the approximated model as it can be seen in Table~\ref{tab:paramters-table 1} that the t-values for the model parameters are larger than the t$_{ref}$. The t values and estimated parameters for MC-MBDoE and DE-MBDoE can also be found in Table~\ref{tab:paramters-table 1-MC}
and Table~\ref{tab:paramters-table 1-disturbance} accordingly.

\begin{table}[H]
    \centering
    \caption{Case study 1: GP-MBDoE. Parameter estimation results in terms of estimated values, 95\% confidence intervals and t-values. }    \begin{tabular}{lcccccc}
    \hline
         Model Parameters & Estimated Values &  95\% t (final) & t$_{ref}$ (final) & 95\% t (initial) & t$_{ref}$(initial) \\
         \hline
        $k_1^0 (\text{min}^{-1})$     & 6.51  & 2.38 & 1.67 & 1.43$^*$ & 1.80\\
        $E_{\text{a}1}  (\text{kJ/mol})$ &  28.3 & 37.28 & 1.67 & 13.04 &1.80\\
        $k_2^0 (\text{min}^{-1})$     & 0.02 &  6.07& 1.67 & 1.15$^*$ &1.80\\
        $E_{\text{a}2}  (\text{kJ/mol})$ & 14.4  &  7.22& 1.67 & 1.93 &1.80\\
        \hline
    \end{tabular}
    \label{tab:paramters-table 1}
\end{table}

\begin{table}[H]
    \centering
    \caption{Case study 1: MC-MBDoE. Parameter estimation results in terms of estimated values, 95\% confidence intervals and t-values. }    \begin{tabular}{lcccccc}
    \hline
         Model Parameters & Estimated Values &  95\% t (final) & t$_{ref}$ (final) & 95\% t (initial) & t$_{ref}$(initial) \\
         \hline
        $k_1^0 (\text{min}^{-1})$     & 6.48  & 6.01 & 1.67& 1.43$^*$ & 1.80\\
        $E_{\text{a}1}  (\text{kJ/mol})$ &  2.83 & 22.01& 1.67 & 13.04 &1.80\\
        $k_2^0 (\text{min}^{-1})$     & 0.02 &  3.09 & 1.67 & 1.15$^*$ &1.80\\
        $E_{\text{a}2}  (\text{kJ/mol})$ & 15.4  &  7.06& 1.67& 1.93 &1.80\\
        \hline
    \end{tabular}
    \label{tab:paramters-table 1-MC}
\end{table}

\begin{table}[H]
    \centering
    \caption{Case study 1: DE-MBDoE. Parameter estimation results in terms of estimated values, 95\% confidence intervals and t-values. }  
    \begin{tabular}{lcccccc}
    \hline
         Model Parameters & Estimated Values &  95\% t (final) & t$_{ref}$ (final) & 95\% t (initial) & t$_{ref}$(initial) \\
         \hline
        $k_1^0 (\text{min}^{-1})$     & 6.88  & 3.86 & 1.67& 1.43$^*$ & 1.80\\
        $E_{\text{a}1}  (\text{kJ/mol})$     & 28.4 & 19.45& 1.67 & 13.04 &1.80\\
        $k_2^0 (\text{min}^{-1})$     & 0.01 &  3.17 & 1.67& 1.15$^*$ &1.80\\
        $E_{\text{a}2}  (\text{kJ/mol})$     & 9.7  & 10.41 & 1.67& 1.93 &1.80\\
        \hline
    \end{tabular}
    \label{tab:paramters-table 1-disturbance}
\end{table}

\begin{figure}
    \centering
    \includegraphics[scale=0.2]{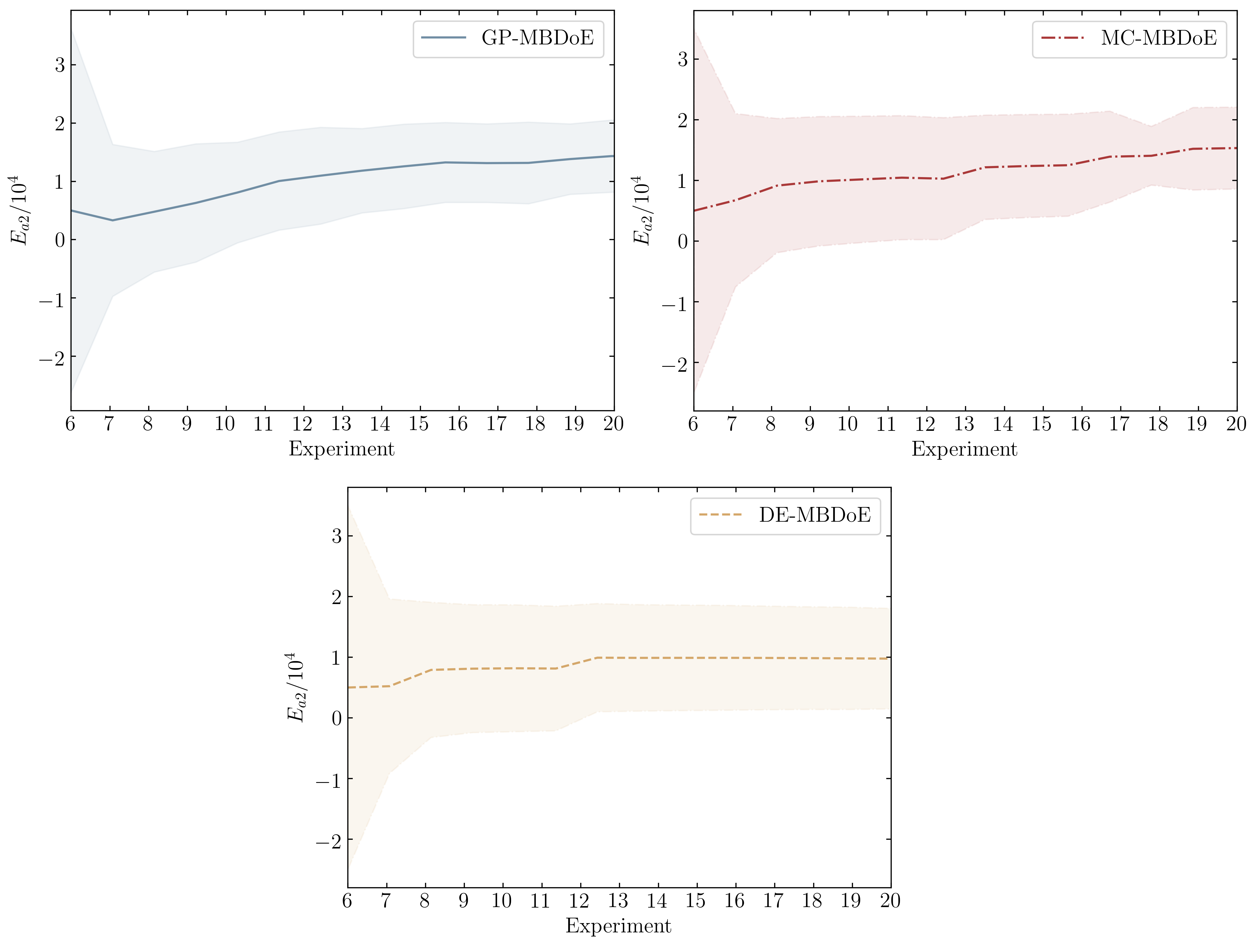}
    \caption{Case Study 1: estimation profiles of $E_{a2}$ for the different designs. The shaded uncertainty area is represented by the square root of the 95\% confidence intervals.}
    \label{fig:theta_CI_simple}
\end{figure}
The parameter are updated in each additive sample that it is obtained. Fig. \ref{fig:theta_CI_simple} presents the trajectory in time for the estimated parameter $E_{a2}$. The behaviour of the parameters is interesting, as their estimation appear to have similar behaviour with the use of MC-MBDoE, which appears to be the least conservative/restrictive (with the most infeasible designs). Notice that due to the model mismatch there are no `true' parameters for the approximated model, that matches the behaviour of the true model. 
%

\begin{table}[H]
    \centering
    \caption{Case study 1: results in terms of chi-square statistics obtained from the different MBDoE configurations.  95\% $\chi^2,$ $\chi^2_{ref} = 74.47$.}  
    \begin{tabular}{lcccccc}
    \hline
         Method & 95\% $\chi^2$ ($\chi^2_{ref} = 74.47$)  \\
         \hline
         GP-MBDoE      & $2018$\\
         MC-MBDoE      & $1965$\\
         DE-MBDoE      & $1888$\\
        \hline
    \end{tabular}
    \label{tab:chi2simple}
\end{table}

Table~\ref{tab:chi2simple} shows that all three methods (i.e. GP-MBDoE, MC-MBDoE, DE-MBDoE) perform similarly in terms of fitting performance. Nevertheless, the GP-MBDoE managed to remain feasible in all the experiments. Interestingly the DE-MBDoE performs better with regards of the $\chi^2$ metric, however the experimental conditions do vary as much as the MC-MBDoE and the GP-MBDoE.

For completeness purposes, we have added the design variables for all the methods (GP-MBDoE, MC-MBDoE, DE-MBDoE) in \ref{sec:DV}.

\subsection{Case Study 2: Flow Reactor: Corrupted Measurements}
\begin{figure}[h]
    \centering
    \includegraphics[scale=0.8]{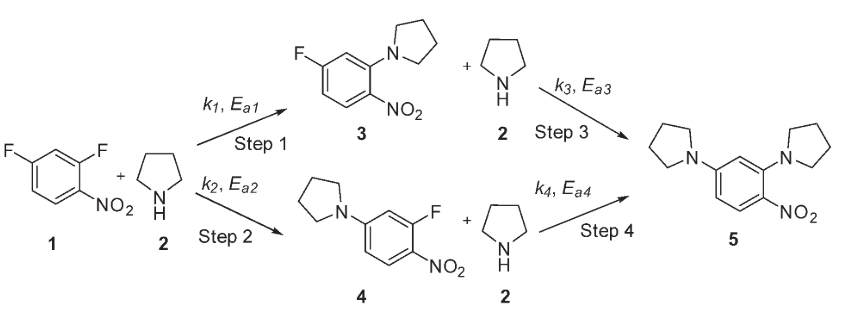}
    \caption{Schematic of the SnAr reaction}
    \label{fig:SNAR}
\end{figure}

To further validate the proposed method, a different case study is presented in the following section, where the model will now be  assumed to be correct but the disturbance will depend on the experimental conditions. 
This second case study considers a nucleophilic aromatic substitution (SNAr) of 2,4-difluoronitrobenzene \{1\} with pyrrolidine \{2\} in ethanol (EtOH) to give a mixture of the desired product ortho-substituted \{3\}, para-substituted \{4\} and bis-product \{5\} as side products. We chose this reaction as a benchmark to evaluate new systems because it produces 3 different major products that are relevant for pharmaceuticals and fine chemicals \cite{Brown2016}.
The inlet flowrates of the reagents 2,4-difluoronitrobenzene \{1\} with pyrrolidine \{2\} ($F_1$ and $F_2$), solvent ethanol ($F_3$) and the temperature $T$  are the design variables: $\textbf{u} = \left[T, F_1, F_2, F_3\right]$
This case study has been adopted from  \cite{Hone2017} and the schematic of the reaction is depicted in Fig.~\ref{fig:SNAR}.
The molar balance of the reactor, using the chemistry from Fig.~\ref{fig:SNAR}, can be written as:
    \begin{equation}
\begin{split}
	&\dfrac{d{ c_1}}{d\tau} =  - r_1 - r_2,~~
	\dfrac{d{c_2}}{d\tau} =  - r_1 - r_2 - r_3 - r_4\\ \\
	&\dfrac{d{ c_3}}{d\tau} =    r_1 + r_3,~~
	\dfrac{d{ c_4}}{d\tau} =    r_2 - r_4,~~
	\dfrac{d{ c_5}}{d\tau} =    r_3 + r_4\\
\end{split}
\end{equation}

with $\tau$ (min) being the residence time, $c_i$ (mol/L) are the molar concentrations for the 2,4-difluoronitrobenzene \{1\}, pyrrolidine \{2\}, ortho-substituted \{3\}, para-substituted \{4\} and bis-product \{5\}.
The reaction rates follow a power law, and the kinetic constant is parametrized as in  Case study 1 (\ref{eq:Arr2}). The values for the kinetic parameters of the true model are adopted from \cite{Hone2017}.

In this case study the assumptions of the kinetic models are correct, however there is a malfunction of an equipment (e.g. HPLC, pumps) that results in a disturbance for the concentration of 2,4-difluoronitrobenzene \{1\}. This systematic error is assumed to be unknown and leads to an overestimation of  concentration measurements of $10\%$ for the 2,4-difluoronitrobenzene \{1\}. An additional concentration measurement noise is considered with zero mean and $10^{-3}$ mol/L standard deviation for all components.  The constraint that is and  considered is $g_1=c_{1}-0.1\leq 0.0$ with probability $\mathbb{P} (g_1\leq0)\geq 0.9$. 

Lower/upper bounds on design variables, defining the design space, are reported in Table \ref{tab:DV_bounds}.
\begin{table}[H]
    \centering
    \caption{Case study 2: definition of design space. }    \begin{tabular}{ccc}
    \hline
         Design Variable & Min &  Max \\
         \hline
        $F_1 \text{(mL/min)}$     & 0.3  & 3.0 \\
        $F_2 \text{(mL/min)}$     & 0.3  & 3.0 \\
        $F_3 \text{(mL/min)}$     & 0.3  & 3.0 \\
        $T(^oC)$                  & 60.0 & 120.\\
        \hline
    \end{tabular}
    \label{tab:DV_bounds}
\end{table}

As in the previous  Latin hypercube is used to generate the experimental conditions for 5 experiments (Table \ref{tab:experiments_2}).

  \begin{table}[H]  
  \centering       
\caption{Case study 2: 5 initial set of experimental conditions for case study 2} \label{tab:experiments_2} 
\begin{tabular}{l|lllll}
	\hline
	 & 1 & 2 & 3 & 4 & 5\\
	\hline
    $T (^oC)$ & 104.8 & 129.9 &  61.1 & 106.9 &
         83.5\\
   $F_1(mL/min)$ &1.0 &   2.5 &   3.0 &   2.4 &
          1.7 \\
    $F_2(mL/min)$ & 1.9  &    2.5 &    0.9   &    1.2 &
          1.6\\
    $F_3(mL/min)$ & 2.1 &   1.9 &      2.7 &      1.3  &  
          1.1\\
	\hline
\end{tabular}
\end{table}

In this case study a D-optimal design is used in all the proposed experimental design configurations. The same parameters for the algorithm used as in case study 1 (Table \ref{tab:opt1}) have been employed. The results follows in the next paragraphs of this section in terms of manipulated inputs, simulated profiles and a posteriori statistics on the final parameter estimation. As in case study 1, all the methodologies are compared with the  same  method  for  propagation  of  uncertainty  in  the  objective  function. Additionally,  the posterior distribution of the parameters is approximated with a normal distribution, which is updated in every iteration. 

\begin{figure}[H]
    \centering
    \includegraphics[scale=0.5]{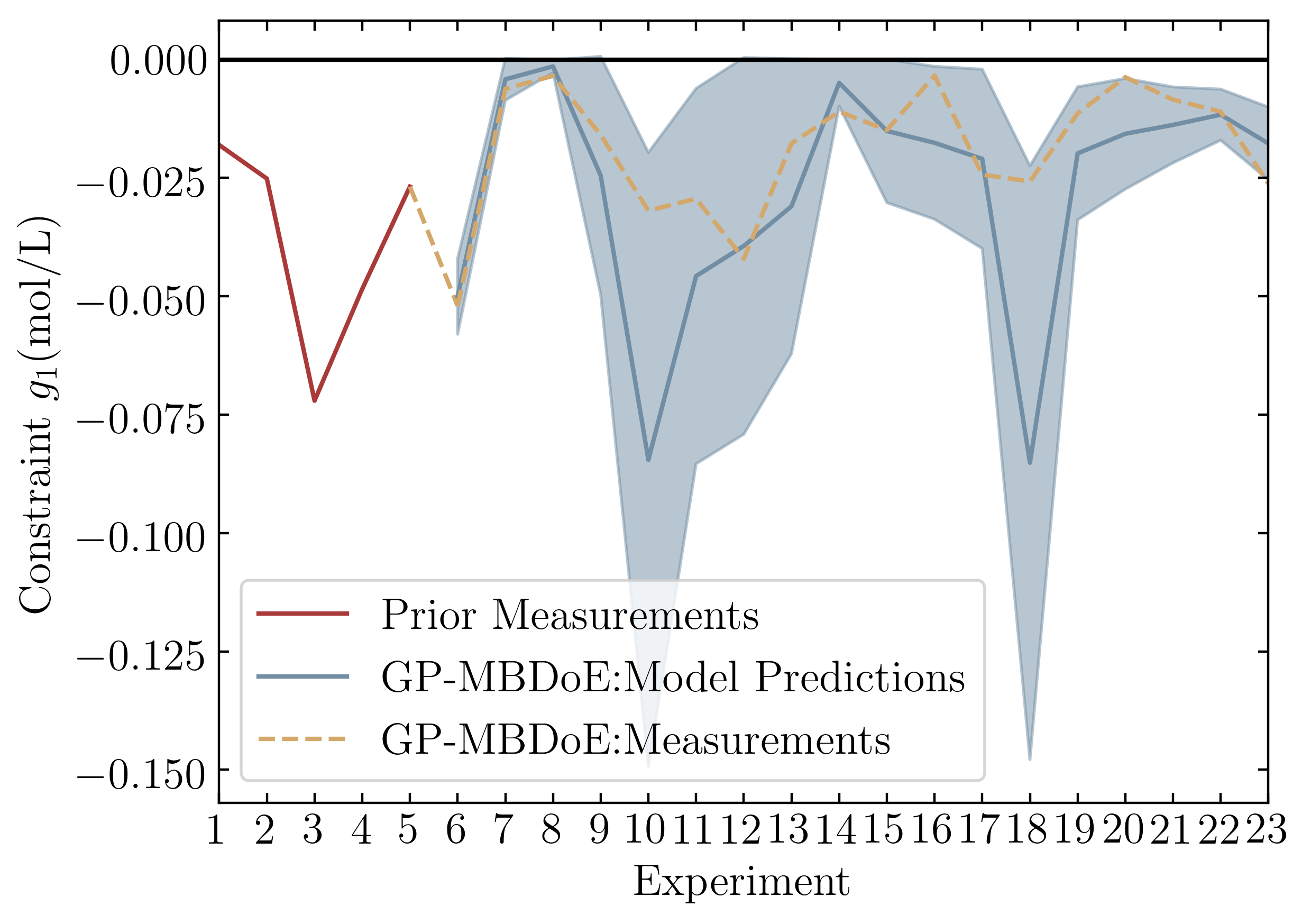}
    \caption{Case study 2: Performance of GP-MBDoE. Blue shaded area represents the 99.7\% confidence interval of the GP. The dashed yellow line is the behaviour of the true model and the red lines represent the value for the constraint at the initial experiments.}
    \label{fig:GP-MBDoE-constraint2}
\end{figure}

\begin{figure}[H]
    \centering
    \includegraphics[scale=0.4]{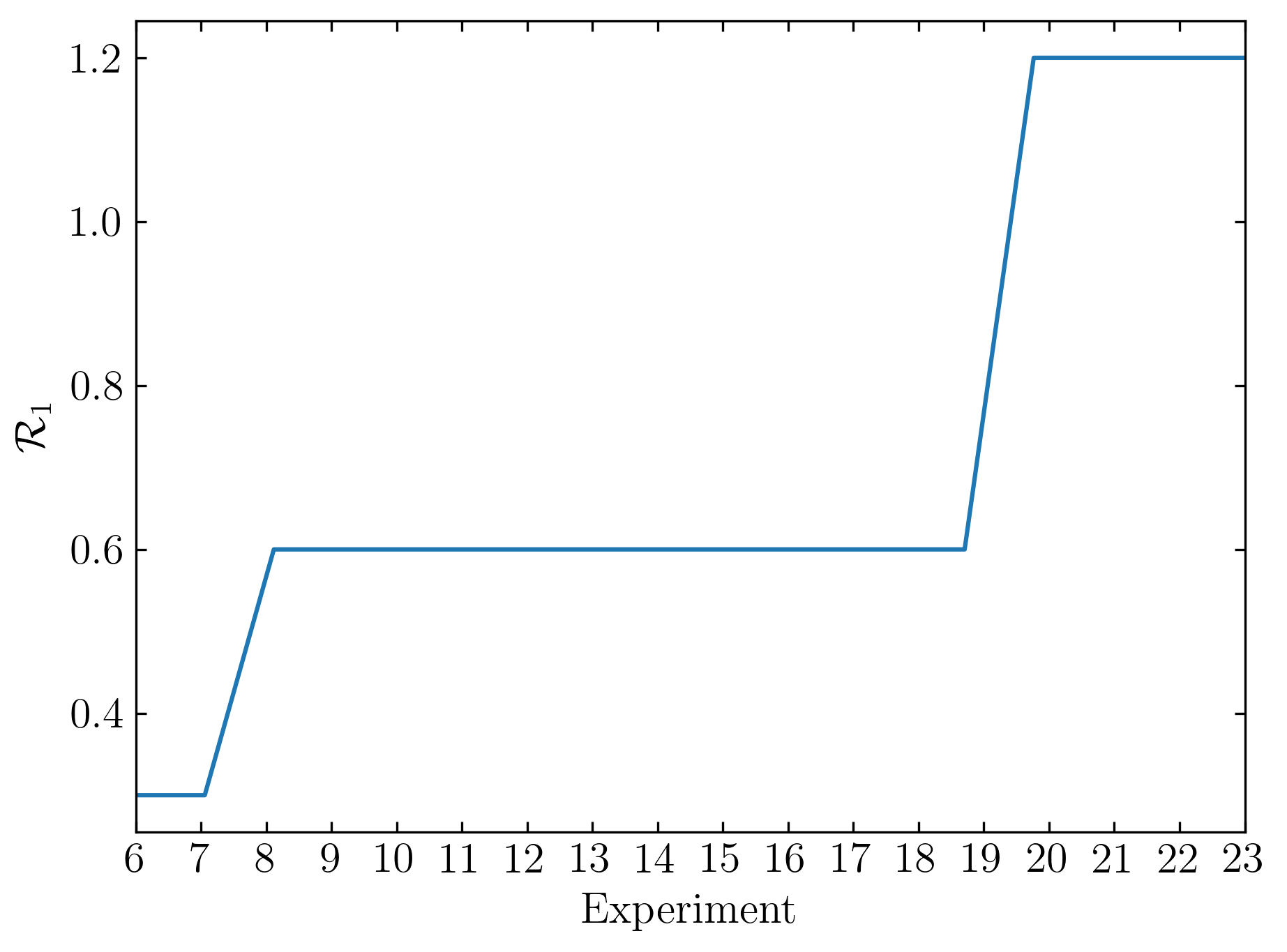}
    \caption{Trust region for the constraint $g_1$}
    \label{fig:TR1}
\end{figure}

All the configurations (MC-MBDoE, DE-MBDoE, GP-MBDoE) are tested in the same online set-up, and the results are presented next.  Fig. \ref{fig:GP-MBDoE-constraint2} depicts the constraint satisfaction of the first constraint $g_1$ for GP-MBDoE. It is evident that the constraint remain feasible in the all the sequence of experiments even though it approaches the bound of the constraint. The blue shaded area represents the 99.7\% confidence intervals of the GP. As expected due to the design space that is used in every consecutive optimization the measurements of the true model lie within the confidence interval of predictions of GP. Notice that the confidence interval of the predictions of the concentration shrinks as more measurements are acquired. 
This behaviour can also be seen in Fig.\ref{fig:TR1}, where the trust region radius $\mathcal{R}_1$ increases as more measurements are acquired. 

\begin{figure}[H]
    \centering
    \includegraphics[scale=0.2]{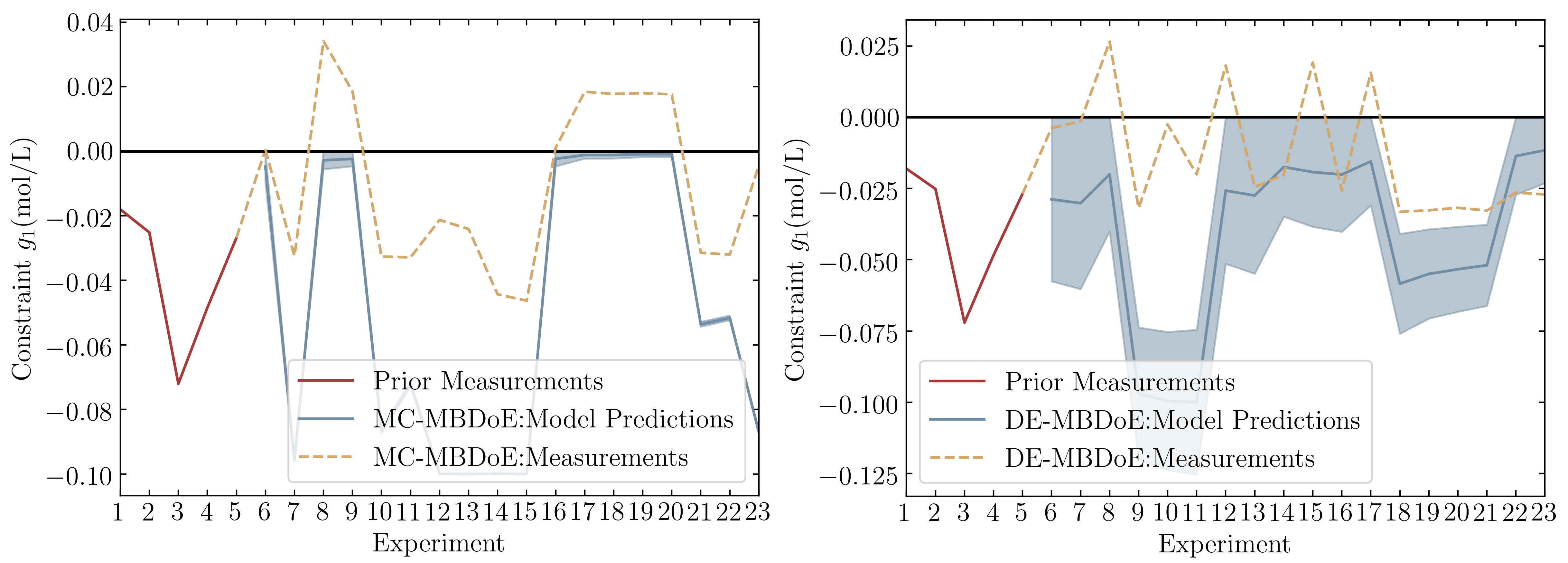}
    \caption{Case study 2: Performance of MC-MBDoE and DE-MBDoE. Blue shaded area represents the computed backoff for each case. The dashed yellow line is the behaviour of the true model and the red lines represent the value for the constraint at the initial experiments.}
    \label{fig:rest-constraint2}
\end{figure}

None of the other formulations managed to remain in the feasible area of the constraint $g_1$, as it is depicted in Fig. \ref{fig:rest-constraint2}. This is not a surprise as none of the aforementioned methods accounted for the structural model mismatch properly. It is noticeable from Fig. \ref{fig:rest-constraint2} that the resulted solution from the optimization is predicted to remain feasible. However, the MC-MBDoE can account only for parametric uncertainty. The posterior of the parameters has been approximated, and the DE-MBDoE can only allow the structural model mismatch to be constant. Additionally, the solution of the MC-MBDoE is also computationally intensive as it requires the use of Monte Carlo sampling from the confidence region of the parameters; in this case, 1000 samples where sampled at each iteration (more details in \cite{Galvanin_backoff}).

Notice that the design with GP-MBDoE defines a test that is now both safe (feasible) and informative for the approximated model, as it can be seen in Table~\ref{tab:paramters-table 2} that the t-values for the model parameters are larger than the t$_{ref}$, which means that they are precisely estimated (very low uncertainty). Nevertheless, high precision does not mean high accuracy, especially in cases with structural mismatch. The t values and estimated parameters for MC-MBDoE and DE-MBDoE can also be found in Table~\ref{tab:paramters-table 2-MC}
and Table~\ref{tab:paramters-table 2-disturbance} accordingly.

\begin{table}[H]
    \centering
    \caption{Case study 2, GP-MBDoE. Parameter estimation results in terms of estimated values, 95\% confidence intervals and t-values.  } 
    \begin{tabular}{lcccccc}
    \hline
         Model Parameters & Estimated Values &  95\% t (final) & t$_{ref}$ (final) & 95\% t (initial) & t$_{ref}$(initial)\\
         \hline
        $k_1^0$(M$^{-1}$min$^{-1}$)     & 5.23  &16.41 & 1.65 & 9.07 & 1.74 \\
        $E_{\text{a}1}  (\text{kJ/mol})$     &  27.4 & 12.17& 1.65 & 6.23 &1.74\\
        $k_2^0$(M$^{-1}$min$^{-1}$)   & 0.33 &  2.92& 1.65 & 1.87 &1.74\\
        $E_{\text{a}2}  (\text{kJ/mol})$ & 33.5  & 2.73 & 1.65 & 0.51$^*$ &1.74 \\
        $k_3^0$(M$^{-1}$min$^{-1}$)    & 0.54  & 18.44 & 1.65 & 4.43 &1.74 \\
        $E_{\text{a}3}  (\text{kJ/mol})$     &  32.8 & 15.91& 1.65 & 6.67 &1.74 \\
        $k_4^0$(M$^{-1}$min$^{-1}$)    & 1.41 &  2.22& 1.65 & 1.28$^*$& 1.74\\
        $E_{\text{a}4} (\text{kJ/mol})$ & 43.6  &  2.62& 1.65 & 9${\times10^{-9}}^*$ & 1.74\\
        \hline
    \end{tabular}
    \label{tab:paramters-table 2}
\end{table}

\begin{table}[H]
    \centering
    \caption{Case study 2, MC-MBDoE. Parameter estimation results in terms of estimated values, 95\% confidence intervals and t-values. }
    \begin{tabular}{lcccccc}
    \hline
         Model Parameters & Estimated Values &  95\% t (final) & t$_{ref}$ (final) & 95\% t (initial) & t$_{ref}$(initial) \\
         \hline
        $k_1^0$(M$^{-1}$min$^{-1}$)     & 4.45 & 28.56 & 1.65 & 9.07 & 1.74 \\
        $E_{\text{a}1} (\text{kJ/mol})$     &  19.70 & 17.44 & 1.65 & 6.23 &1.74\\
        $k_2^0$(M$^{-1}$min$^{-1}$)    & 0.28 &  3.95 & 1.65 & 1.87 &1.74\\
        $E_{\text{a}2} (\text{kJ/mol})$ & 27.6  &  3.35 & 1.65 & 0.51$^*$ &1.74 \\
        $k_3^0$(M$^{-1}$min$^{-1}$)    & 0.47  & 16.54 & 1.65 & 4.43 &1.74 \\
        $E_{\text{a}3} (\text{kJ/mol})$     &  33.0 & 15.97 & 1.65 & 6.67 &1.74 \\
        $k_4^0$(M$^{-1}$min$^{-1}$)    & 1.32 &  1.8 & 1.65 & 1.28$^*$& 1.74\\
        $E_{\text{a}4} (\text{kJ/mol})$ & 43.8  & 1.94 & 1.65 & 9${\times10^{-9}}^*$ & 1.74\\
        \hline
    \end{tabular}
    \label{tab:paramters-table 2-MC}
\end{table}

\begin{table}[H]
    \centering
    \caption{Case study 2, DE-MBDoE. Parameter estimation results in terms of estimated values, 95\% confidence intervals and t-values. }
    \begin{tabular}{lcccccc}
    \hline
         Model Parameters & Estimated Values &  95\% t (final) & t$_{ref}$ (final) & 95\% t (initial) & t$_{ref}$(initial) \\
         \hline
        $k_1^0$(M$^{-1}$min$^{-1}$)   & 4.31  & 23.71 & 1.65 & 9.07 & 1.74 \\
        $E_{\text{a}1} (\text{kJ/mol})$     &  24.2 & 17.16& 1.65 & 6.23 &1.74\\
        $k_2^0$(M$^{-1}$min$^{-1}$)    & 0.33 &  2.45& 1.65 & 1.87 &1.74\\
        $E_{\text{a}2} (\text{kJ/mol})$ & 27.6  &  2.11 & 1.65 & 0.51$^*$ &1.74 \\
        $k_3^0$(M$^{-1}$min$^{-1}$)    & 0.42& 4.22 & 1.65 & 4.43 &1.74 \\
        $E_{\text{a}3} (\text{kJ/mol})$     &  39.9 & 5.31& 1.65 & 6.67 &1.74 \\
        $k_4^0$(M$^{-1}$min$^{-1}$)   & 1.98 &  1.327$^*$& 1.65 & 1.28$^*$& 1.74\\
        $E_{\text{a}4} (\text{kJ/mol})$ & 34.3  &  1.44$^*$ & 1.65 & 9${\times10^{-9}}^*$ & 1.74\\
        \hline
    \end{tabular}
    \label{tab:paramters-table 2-disturbance}
\end{table}

\begin{figure}[H]
    \centering
    \includegraphics[scale=0.2]{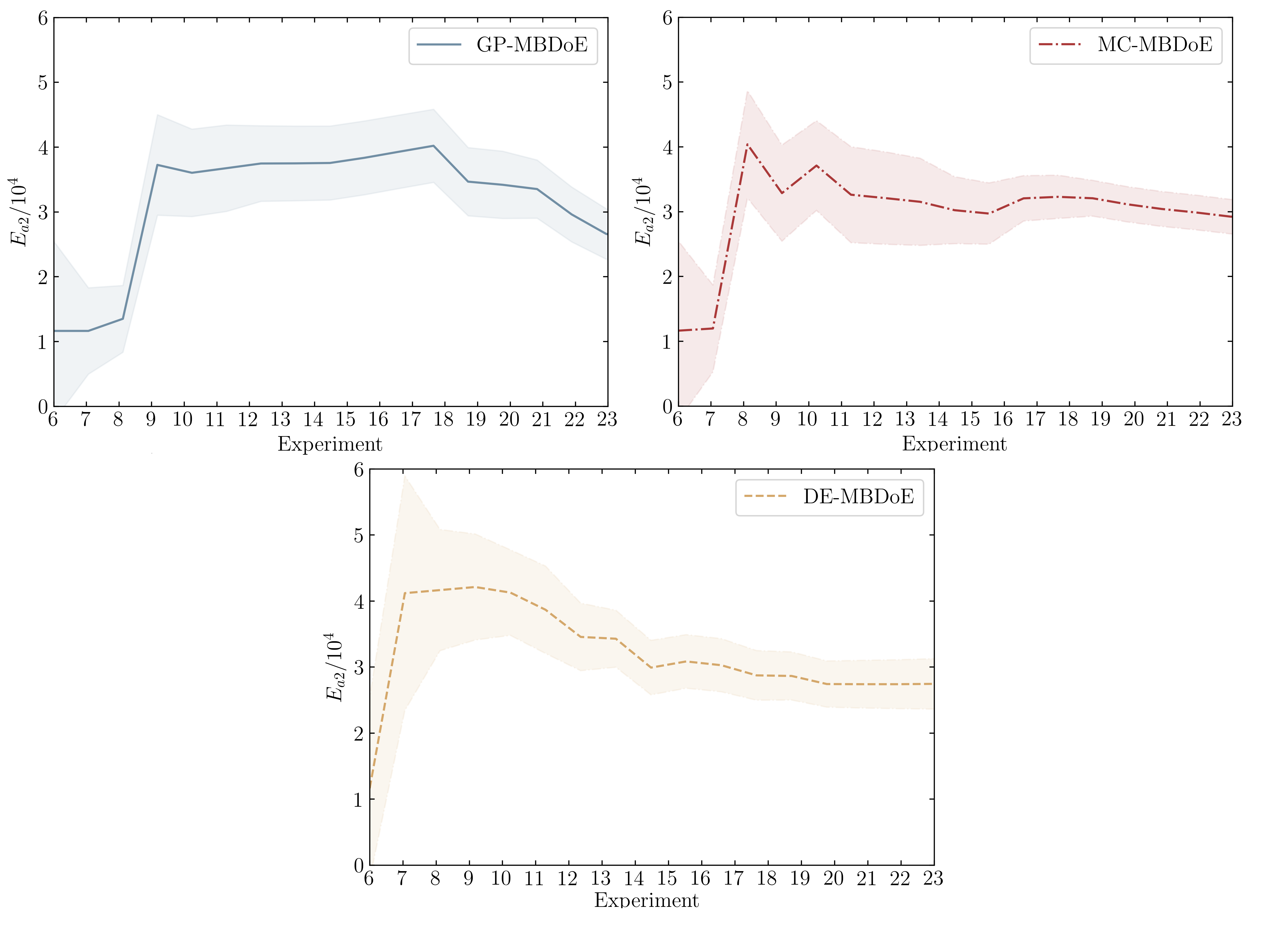}
    \caption{Case Study 2: estimation profiles of $E_{a2}$ for the different designs. The shaded uncertainty area is represented by the square root of the 95\% confidence intervals.}
    \label{fig:theta_CI_simple2}
\end{figure}
As in the previous case, the parameters are updated after each new sample is acquired from experiments. The parameter $E_{a2}$ is illustrated in Fig. \ref{fig:theta_CI_simple2} and initially does not pass the $t$-test. The behaviour of the parameters for each design appears to be similar, with the MC-MBDoE and  DE-MBDoE  having a smoother behaviour of the estimates as the parametric uncertainty reduces. Nevertheless, the parameter estimates are not asymptotically stable for MC-MBDoE and GP-MBDoE. This behaviour could be potentially be caused by the structural mismatch that is present in the models. 
%

The result here is similar with Case study 1, where the proposed GP-MBDoE is performs better the MC-MBDoE, but with feasible designs.

\begin{table}[H]
    \centering
    \caption{Case study 2, 95\% $\chi^2,$ $\chi^2_{ref} = 132.14$.}  
    \begin{tabular}{lcccccc}
    \hline
         Method & $\chi^2$  \\
         \hline
         GP-MBDoE      & $1805$\\
         MC-MBDoE      & $3892$\\
         DE-MBDoE      & $3079$\\
        \hline
    \end{tabular}
    \label{tab:chi2complex}
\end{table}

In this case study the Table~\ref{tab:chi2complex} shows that the GP-MBDoE managed to outperform the standard robust MBDoE methods (MC-MBDoE and DE-MBDoE), performing better in terms of their predicting capabilities. Additionally, the GP-MBDoE managed to remain feasible in all the experiments.

For completeness purposes, we have added the design variables for all the methods (GP-MBDoE, MC-MBDoE, DE-MBDoE) in \ref{sec:DV}.

\section{Discussion}\label{sec:Discussion}

In this section, an additional discussion regarding the results of the case studies is added. In the case studies, theour proposed method (GP-MBDoE) was compared with two alternative methods, a method that deals with parametric only uncertainty (MC-MBDoE) and a disturbance based estimation method (DE-MBDoE). The GP-MBDoE has the main advantage: it can deal with potential model-mismatch and have as good performance as the standard MBDoE techniques in terms of information metrics.
The non-parametric nature of the GPs offers the opportunity to approximate the physical system with a small number of data-points locally with the help of trust regions. This nice property comes with additional experiments' expense due to the conservative nature of the trust regions. This method's main computational burden is the training of the second Gaussian process for the objective function. An arbitrary large amount of data-points can be used for this surrogate, as only in-silico data points are needed. This can be an issue as the number of the design variables and model parameters increases; however, approximated methods have been proposed, including sparse GPs and generally variational methods for the training  \cite{pmlr-v5-titsias09a}. In this work, the computational times needed are on average 42 and 121 CPU-sec with 10 multi-starts for each GP and 10 multi-starts for the optimization for case study 1 and case study 2, respectively. The computational times for the GP-MBDoE, MC-MBDoE and DE-MBDoE are given in Table~\ref{tab:cpu cost}. 
Please note that a preliminary set of experiments is required, particularly we need $n_u+1$ data points to compute the initial Gaussian process as the covariance of the input data points needs to be invertible.

The MC-MBDoE method, on the other hand, can effectively deal with parametric uncertainty and has proven to be quite effective \cite{Telen2018, Nimmegeers2020}.  Nevertheless, even in the absence of model mismatch, such an approach may have several issues:
Despite the simplicity of such approaches, neither robust optimization nor stochastic programming addresses the ambiguity of the uncertainty set, i.e. the issue is to have a reliable representation of the parametric uncertainty region and the uncertainty for the confidence region itself.
This issue could be solved by applying 1) an initial conservative design to account for small number of data and then 2)  the standard MBDoE techniques in a receding horizon fashion \cite{Galvanin2012_DIST} and updating the uncertainty set. In this paper, this method's computational time is high due to the Monte Carlo approach that is followed, specifically 250 CPU-sec and 500 CPU-sec. Nevertheless, as mentioned, more computationally efficient methods have been proposed in the literature to solve optimal control problems with parametric uncertainty.

The DE-MBDoE, on the other hand, can be seen as a subcategory of the GP-MBDoE (as a constant value disturbance model is used) without the trust regions. The DE-MBDoE effectively can approximate the model mismatch with a constant value similarly to the well-known disturbance rejection techniques \cite{MPC_dist}. This method can be very effective mainly when the DE-MBDoE is applied in a receding horizon fashion as the constant value is updated very often. The disadvantages here are 1) the constant value may not represent the model mismatch and 2) the method may be over-confident, where regions far from the current point may be explored. This method is the most computationally efficient, as it does not require any MC, resulting in less than 20 CPU-sec for the first case study and less than 62 CPU-sec for the second case study.

\begin{table}[H]
    \centering
    \caption{Computational times for each case study and method}  
    \begin{tabular}{lcccccc}
    \hline
         Method &
         $\begin{array}{c}
             \text{Case Study 1:} \\
              \text{CPU time(sec)}
         \end{array}$ &
         $\begin{array}{c}
             \text{Case Study 2:} \\
              \text{CPU time(sec)}
         \end{array}$  \\
         \hline
         GP-MBDoE      & $42$ & $121$\\
         MC-MBDoE      & $250$ & $500$\\
         DE-MBDoE      & $20$ & $62$\\
        \hline
    \end{tabular}
    \label{tab:cpu cost}
\end{table}

\section{Conclusions}
In this article a new methodology for the safe MBDoE based on GP and trust regions in the presence of structural uncertainty. GP-MBDoE was proposed and discussed; The optimal design of an experiment for improving parameter estimation is a particular form of optimization problem that can be very effective, but both optimality and feasibility of the designed experiment are important issues to consider. The proposed strategy allows to restrict the feasible space maintaining the optimal design in the feasible (safe) region of the state variables. Two simulated (in-silico) case studies have been proposed to assess the effectiveness of the new technique. In the first case study, the methodology was applied to a
model of of a chemical reaction in a flow reactor, where the assumed mechanism is inadequate to represent the system. The constraint tightening strategy allows estimating the parametric set describing the approximate model as well as run the experiments in the safe feasible region.  In the second case study, dealing with a larger problem of chemical parallel reactions (SnAr), where the model assumptions are correct, however there is a presence of a disturbance that is dependent on the design variables due to equipment malfunction. The proposed methodology (GP-MBDoE) managed to offer predictive capabilities similar (Case study 1) or better (Case study 2) compared to the standard robust MBDoE methods (MC-MBDoE \& DE-MBDoE), whilst resulting in feasible experimental designs.
Future work will aim at further improving the proposed
approach by using methods related to Bayesian calibration to calibrate the approximated model itself \cite{Kennedy2001} via a Bayesian framework. The parameter of the kinetic model can be seen as calibration parameters and estimated through their posterior distribution, which provides a non-asymptotic approach for uncertainty quantification. Additionally, the structural mismatch can be potentially be diagnosed \cite{QUAGLIO2020106659} and different structure could be potential be found using artificial neural networks \cite{QUAGLIO2020106759}.

\section*{Acknowledgment}
FThis project has received funding from the EPSRC grantprojects (EP/R032807/1) is gratefully acknowledged.

\bibliography{references}

\clearpage

\appendix

\section{Uncertainty propagation with Gaussian processes}\label{sec:Uncertainty propagation}

GP has found a success in the literature due to its attribute to estimate not only the expected value of a function (mean) but also the variance. However, the expressions for the posterior distribution of $f_i$ in (\ref{eq:posterior}) hold only for a deterministic input ${\textbf{x}^*}$. In the case that ${\textbf{x}^*}$ is random variable, then the expression (\ref{eq:posterior}) do not hold and for the case that ${\textbf{x}^*}\sim \mathcal{N}\big( \pmb{\mu}_{{\textbf{x}^*}},\pmb{\Sigma}_{{\textbf{x}^*}}\big)$ the posterior is :
\begin{equation}
    p\big(f_i({\textbf{x}^*})|\pmb{\mu}_{\textbf{x}^*},\pmb{\Sigma}_{\textbf{x}^*},\mathcal{D}_i\big) = \int  p\big(f_i({\textbf{x}^*})|{\textbf{x}^*},\mathcal{D}_i\big)p\big({\textbf{x}^*}\big)d{\textbf{x}^*},
\end{equation}
which is computationally intractable.
\subsection{Gaussian Approximation: Formulation}\label{sec:Gaussian_approxim}
To accommodate this issue, an analytical Gaussian approximation has been found, where  only the mean ($\hat{m}_i(\pmb{\mu}_{{\textbf{x}^*}},\pmb{\Sigma}_{{\textbf{x}^*}})$) and variance ($\hat{\Sigma}_i(\pmb{\mu}_{{\textbf{x}^*}},\pmb{\Sigma}_{{\textbf{x}^*}})$) of $f_i({\textbf{x}^*})|\pmb{\mu}_{\textbf{x}^*},\pmb{\Sigma}_{{\textbf{x}^*}},\mathcal{D}_i$ is computed. These are obtained employing the law of iterated expectations and law of conditional variances \cite{Girard2002}
\begin{equation}\label{eq:gaussian_aprx}
    \begin{split}
       \hat{m}_i(\pmb{\mu}_{\textbf{x}^*},\pmb{\Sigma}_{{\textbf{x}^*}}) &=  \mathbb{E}_{\textbf{x}^*}\Big(\mathbb{E}_{{f}_i^*}\big( f_i^*(\textbf{x}^*)|{\textbf{x}^*}\big) \Big) = \mathbb{E}_{\textbf{x}^*}\Big(m_i({\textbf{x}^*}) \Big)\\
       \hat{\Sigma}_i(\pmb{\mu}_{\textbf{x}^*},\pmb{\Sigma}_{{\textbf{x}^*}}) &= \mathbb{E}_{\textbf{x}^*}\Big(\mathbb{V}_{{f}_i^*}\big( f_i^*(\textbf{x}^*)|{\textbf{x}^*}\big) \Big) + \mathbb{V}_{\textbf{x}^*}\Big(\mathbb{E}_{{f}_i^*}\big( f_i^*(\textbf{x}^*)|{\textbf{x}^*}\big) \Big)  \\
        &=\mathbb{E}_{\textbf{x}^*}\Big(\Sigma_i(\textbf{x}^*)) \Big) + \mathbb{V}_{\textbf{x}^*}\Big(m_i({\textbf{x}^*}) \Big).
    \end{split}
\end{equation}
The expressions in~(\ref{eq:gaussian_aprx}) can analytically be found for the case of squared exponential. In the case of general case of kernels, Taylor expansion has been proposed to find a close form expression of (\ref{eq:gaussian_aprx}). 
First order order Taylor expansion is applied around $\pmb{\mu}_{\textbf{x}^*}$ for ${m}_i(\cdot,\cdot)$:
\begin{equation}\label{eq:taylor_mean}
\begin{split}
 &\mathbb{E}_{\textbf{x}^*}\Big(m_i({\textbf{x}^*}) \Big) \approx \mathbb{E}_{\textbf{x}^*}\Big(m_i(\pmb{\mu}_{\textbf{x}^*}) +\left.\dfrac{\partial {m}_i(\textbf{x}^*) }{\partial \textbf{x}^*}\right\vert^T_{\pmb{\mu}_{\textbf{x}^*}}(\textbf{x}^* - \pmb{\mu}_{\textbf{x}^*})\Big)=m_i(\pmb{\mu}_{\textbf{x}^*})\\
 &\mathbb{V}_{\textbf{x}^*}\Big(m_i({\textbf{x}^*}) \Big) \approx
 \mathbb{V}_{\textbf{x}^*}\Big(m_i(\pmb{\mu}_{\textbf{x}^*}) +\left.\dfrac{\partial {m}_i(\textbf{x}^*) }{\partial \textbf{x}^*}\right\vert^T_{\pmb{\mu}_{\textbf{x}^*}}(\textbf{x}^* - \pmb{\mu}_{\textbf{x}^*})\Big)=
\left.\dfrac{\partial {m}_i(\textbf{x}^*) }{\partial \textbf{x}^*}\right\vert^T_{\pmb{\mu}_{\textbf{x}^*}}\pmb{\Sigma}_{\textbf{x}^*}\left.\dfrac{\partial {m}_i(\textbf{x}^*) }{\partial \textbf{x}^*}\right\vert_{\pmb{\mu}_{\textbf{x}^*}}
\end{split}
\end{equation}
and second order Taylor for the variance ${\Sigma}_i(\cdot,\cdot)$:
\begin{equation}\label{eq:taylor_variance}
    \begin{split}
    \Sigma_i({\textbf{x}^*}) &\approx \Sigma_i(\pmb{\mu}_{\textbf{x}^*}) +\left.\dfrac{\partial {\Sigma}_i(\textbf{x}^*) }{\partial \textbf{x}^*}\right\vert^T_{\pmb{\mu}_{\textbf{x}^*}}(\textbf{x}^* - \pmb{\mu}_{\textbf{x}^*}) +\dfrac{1}{2}(\textbf{x}^* - \pmb{\mu}_{\textbf{x}^*})^T\left.\dfrac{\partial^2 {\Sigma}_i(\textbf{x}^*) }{\partial \textbf{x}^*\partial {\textbf{x}^*}^T}\right\vert_{\pmb{\mu}_{\textbf{x}^*}}(\textbf{x}^* - \pmb{\mu}_{\textbf{x}^*})\\
        \mathbb{E}_{\textbf{x}^*}\Big(\Sigma_i({\textbf{x}^*}) \Big) &\approx
        \Sigma_i(\pmb{\mu}_\textbf{x}^*) + \dfrac{1}{2}\text{trace}\Big\{\left.\dfrac{\partial^2 {\Sigma}_i(\textbf{x}^*) }{\partial \textbf{x}^*\partial {\textbf{x}^*}^T}\right\vert_{\pmb{\mu}_{\textbf{x}^*}}\pmb{\Sigma}_{\textbf{x}^*} \Big\}.
    \end{split}
\end{equation}
Now, the approximated mean $\hat{m}_i$ and variance $\hat{\Sigma}_i$ can be computed by substituting (\ref{eq:taylor_mean})\&(\ref{eq:taylor_variance}) to (\ref{eq:gaussian_aprx}):
\begin{equation}\label{eq:gaussian_aprx_forms}
    \begin{split}
       &\hat{m}_i(\pmb{\mu}_{\textbf{x}^*},\pmb{\Sigma}_{{\textbf{x}^*}}) = m_i(\pmb{\mu}_{\textbf{x}^*})\\
       &\hat{\Sigma}_i(\pmb{\mu}_{\textbf{x}^*},\pmb{\Sigma}_{{\textbf{x}^*}}) =\Sigma_i(\pmb{\mu}_\textbf{x}^*) + \dfrac{1}{2}\text{trace}\Big\{\left.\dfrac{\partial^2 {\Sigma}_i(\textbf{x}^*) }{\partial \textbf{x}^*\partial {\textbf{x}^*}^T}\right\vert_{\pmb{\mu}_{\textbf{x}^*}}\pmb{\Sigma}_{\textbf{x}^*} \Big\} + \left.\dfrac{\partial {m}_i(\textbf{x}^*) }{\partial \textbf{x}^*}\right\vert^T_{\pmb{\mu}_{\textbf{x}^*}}\pmb{\Sigma}_{\textbf{x}^*}\left.\dfrac{\partial {m}_i(\textbf{x}^*) }{\partial \textbf{x}^*}\right\vert_{\pmb{\mu}_{\textbf{x}^*}}.
    \end{split}
\end{equation}
The approximated solution of the Gaussian approximation can be advantageous for the the propagation of uncertainty. In this paper  approximation (\ref{eq:gaussian_aprx_forms}) will be used to propagate the uncertainty of the model parameters to the Fisher information matrix.

\subsection{Gaussian Approximation: Simple Numerical Example}\label{sec:Gaussian_approxim_example}

\begin{figure}[H]
    \centering
    \includegraphics[scale=0.5]{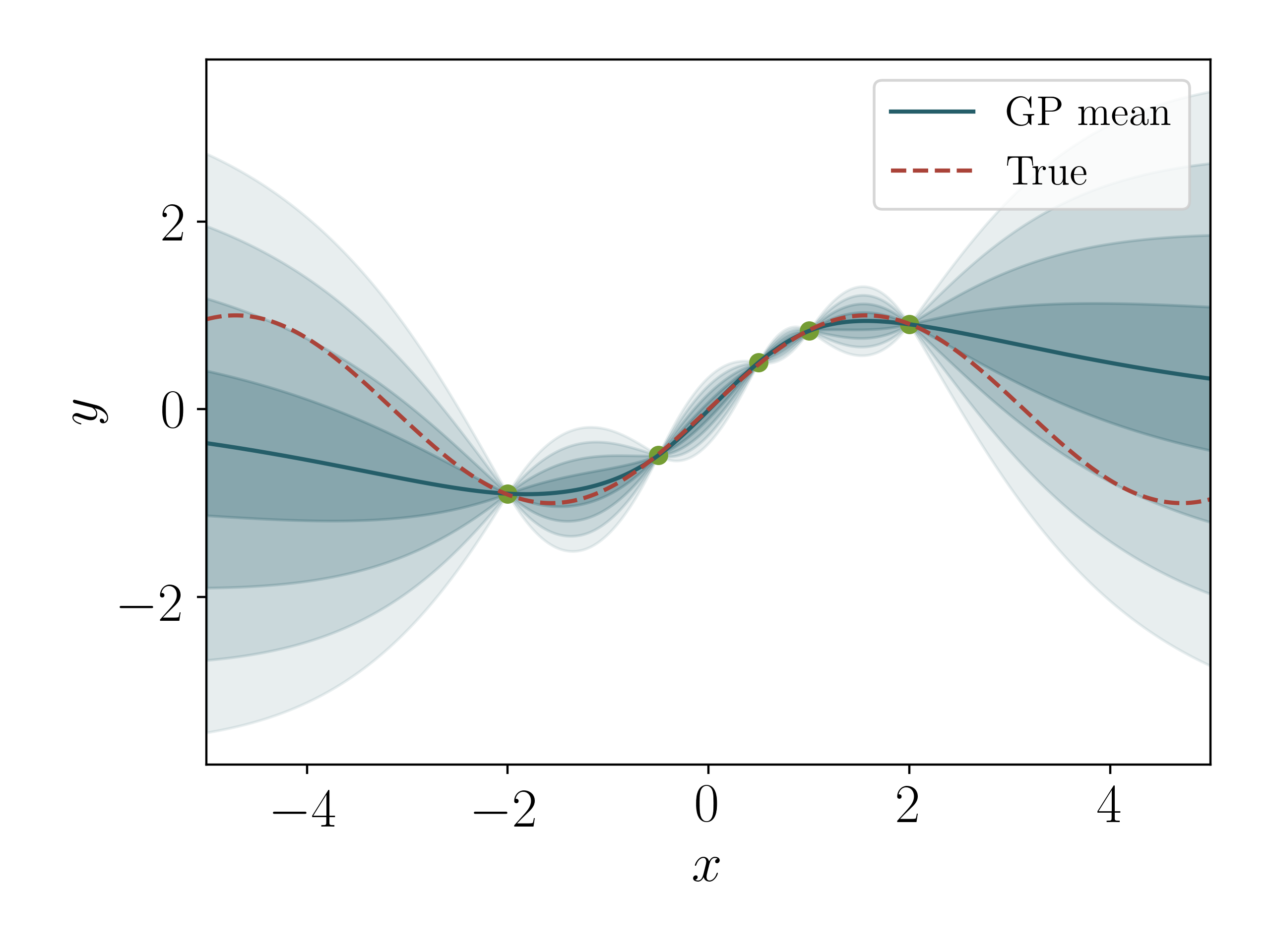}
    \caption{The ground (red dashed line) is the function that the GP approximates given the available data. The GP predictions for the mean ($\mu$) is the blue line with variance $\Sigma$, and the shaded areas represents the area between $m\pm\sqrt{\Sigma}, m\pm\sqrt{\Sigma}, \mu\pm3\sqrt{\Sigma}, m\pm4\sqrt{\Sigma}$ and  $m\pm10\sqrt{\Sigma}$.}
    \label{fig:simple_GP_sin}
\end{figure}

Let a simple function $f(x) = \sin(x)$ be assumed as the `true' model, with $x\in \mathbb{R}$ and the collected data are corrupted by a Gaussian noise $\epsilon_f$ with standard deviation 0.01, $y=f(x) + \epsilon_f$. A GP is used to approximate the data points, specifically a squared-exponential (SE) covariance function is used and its hyperparameters are fitted using maximum likelihood estimation (with multi-starts  to avoid local optima). The mean $m$ and variance $\Sigma$ of the GP for each $x$ are illustrated in Fig~\ref{fig:simple_GP_sin}. 

\begin{figure}[H]
    \centering
    \includegraphics[scale=0.5]{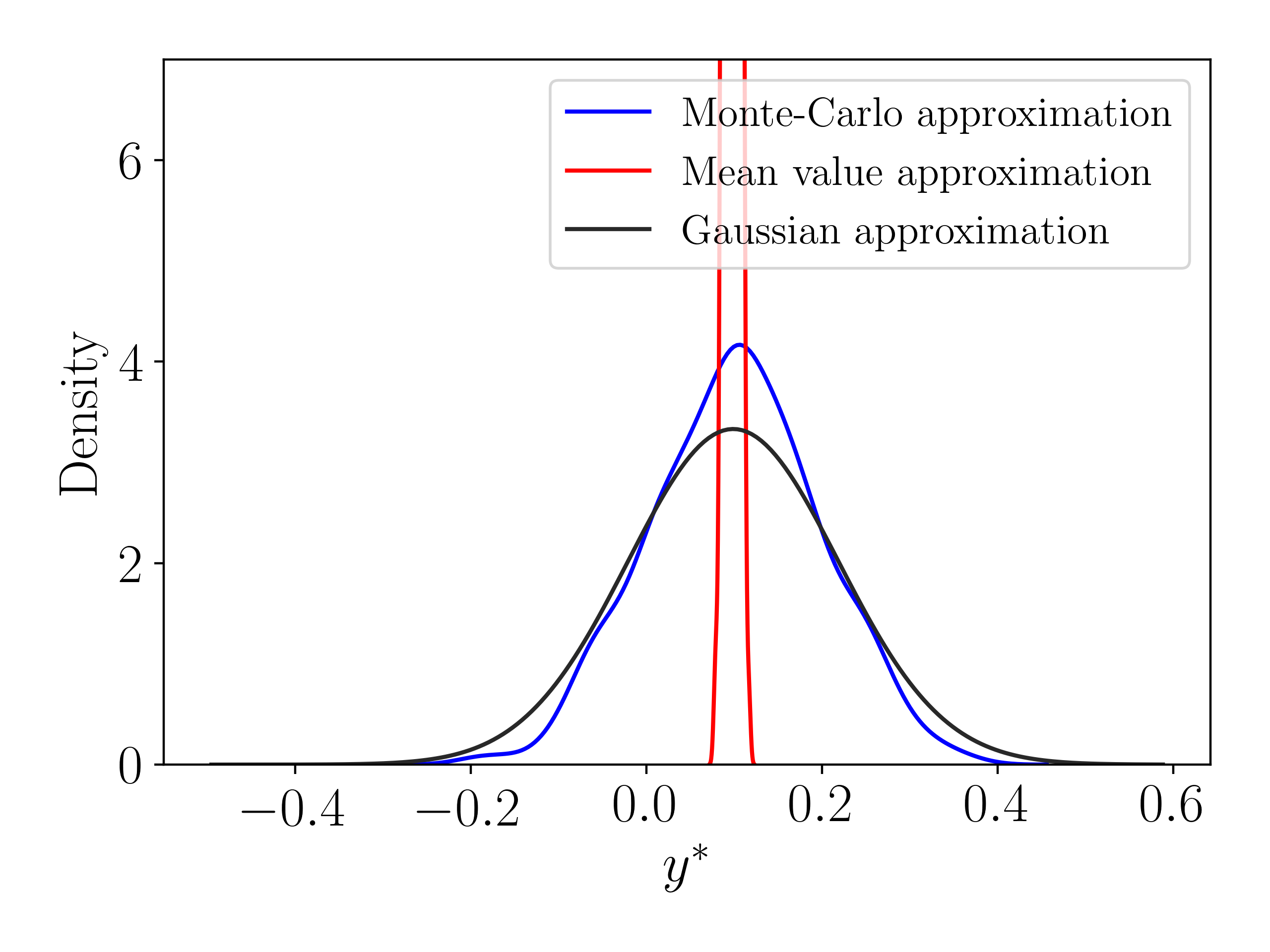}
    \caption{Comparison of prediction methods for random input $x^*$. The posterior distribution is evaluated with the different approximation techniques; Monte-Carlo approximation ({\color{blue} blue}), Mean value approximation ({\color{red} red}) and Gaussian approximation ({\color{black} black}).}
    \label{fig:gaussian_approximation_example_density}
\end{figure}

The shaded areas show the $m+r \sqrt{\Sigma}$ for $r=1, 2, 3, 4$ and $10$. If the test point $x^*$ is deterministic then the prediction for $f(x^*)$ would follow a Gaussian distribution with mean $m$ and variance $\Sigma$; however such computation is not correct if $x^*$ is a random variable. Let's assume that $x^*\sim \mathcal{N}\big( 0.1,0.01\big)$ then Fig. \ref{fig:gaussian_approximation_example_density} depicts 3 different approaches of computing the posterior density function; 1) Monte-Carlo approximation: This distribution is approximated using Monte-Carlo sampling (1000 samples), 2) Mean value approximation: The computation of the posterior ignores the stochastic nature of the test point $x^*$ and computes the posterior $m$ and $\Sigma$ using (\ref{eq:posterior}), and 3) Gaussian approximation: The posterior distribution is approximated using the analytical Gaussian approximation (\ref{eq:gaussian_aprx_forms}). 

It is apparent that the posterior distribution is not Gaussian and that the approximation as a Gaussian distribution leads to error, even if the first mean and variance are exactly the same. The Mean value Approximation can be very conservative or over-confident (like in this example) as it neglects into account the stochastic nature of the input.

\newpage

\section{Nomenclature}
\begin{table}[h]
  \centering
  \caption{Nomenclature}
    \begin{tabular}{cl}
    \hline
    Symbol & Description \\
    \hline
    $\textbf{x}$     & States \\
    $\textbf{u}$     & Design variables \\
    $\textbf{w}$     & External disturbances \\
    $\pmb{\nu}$ & Random noise \\
        $\hat{\pmb{\theta}}$ & Model parameters \\
    $\pmb{\theta}$ & Model parameters \\
    $\hat{(\cdot)}$ & Approximate quantity $(\cdot)$\\
    $n_x$ & Dimension of $\textbf{x}$\\
    $\textbf{V}_{\pmb{\theta}}$ & Variance-covariance of the parameters $\pmb{\theta}$ \\
    $\mathcal{D}$     & Data available \\
    $\mathcal{L}(\pmb{\theta}|\mathcal{D})$ & The joint probability of the prediction-observation mismatch in all data points ($\mathcal{D}$) for the parameter $\pmb{\theta}$\\
    $\mathbb{E}$     & Expectation \\
    $\mathbb{P}$      & Probability \\
    $\mathbb{V}$      & Variance  \\
    $\mathbb{U}$      & Constraints for the design variables \\
    $\dot{\textbf{x}}$ & derivative with respect to time of variable $\textbf{x}$ \\
    $g_j$  & Constraint $j$ \\
    $\mathcal{G}\mathcal{P}$    & Gaussian Process (Distribution over functions)\\
        \hline
    \end{tabular}%
  \label{tab:addlabel}%
\end{table}%
\begin{table}[h]
  \centering
  \caption{Nomenclature (continued)}
    \begin{tabular}{cl}
    \hline
    Symbol & Description \\
    \hline
    $\mu_i$ & Prior mean for GP \\
    $k_i$  & Kernel function for GP of $i^{th}$ state\\
    $K$ or $K(\textbf{X},\textbf{X})$ & Covariance (or Gram) matrix\\
    $m_i$  & Posterior mean for GP of $i^{th}$ state\\
    $\Sigma_i$ & Variance of $i^{th}$ state \\
    $\mathcal{N}(\cdot,\cdot)$ & Normal distribution defined by two quantities, i.e. mean and variance-covariance \\
    $\sim$ & Distributed according to; example: $x \sim \mathcal{N}(\mu, \sigma^2) $\\
    $||\cdot||_{k_i}$ & Torm in the associated Reproducing Kernel Hilbert Space\\
    $\mathcal{H}_k$ & Reproducing kernel Hilbert space (RKHS)\\
    $\mathcal{R}$  & Radius of trust region \\
    $\rho$ & Percentage error of model/Measurement \\
    $t_1$,$t_2$ & Adaptation parameters of trust region \\
    $\eta_1$, $\eta_2$ & Threshold for $rho$ \\
    $t$     & $t$ value for $t$-test \\
    $\chi^2$ & Chi square value  \\
    $\bar{\alpha}$ & Probability of constraints violation \\
    $ 1-\alpha$ & Confidence level\\ 
    $\alpha_J$ & Exploration term for MBDoE \\
    $k_j^r$ & Kinetic reaction parameter of $j^{th}$ reaction \\
    $k_j^0$ & Pre-exponential of $j^{th}$ reaction \\
    $E_{a_j}$ & Activation energy of $j^{th}$ reaction\\
    $F$     & Flowrate \\
    $c$     & Concentrations \\
    $T$     & Temperature \\
    \hline
    \end{tabular}%
  \label{tab:addlabel2}%
\end{table}%
\newpage
\section{Design variables for Case studies}\label{sec:DV}

In this section the design variables for the case study 1 and 2 are given. 
\subsection{Case study 1: Design Variables}\label{DV1}
\begin{figure}[H]
    \centering
    \includegraphics[scale=0.2]{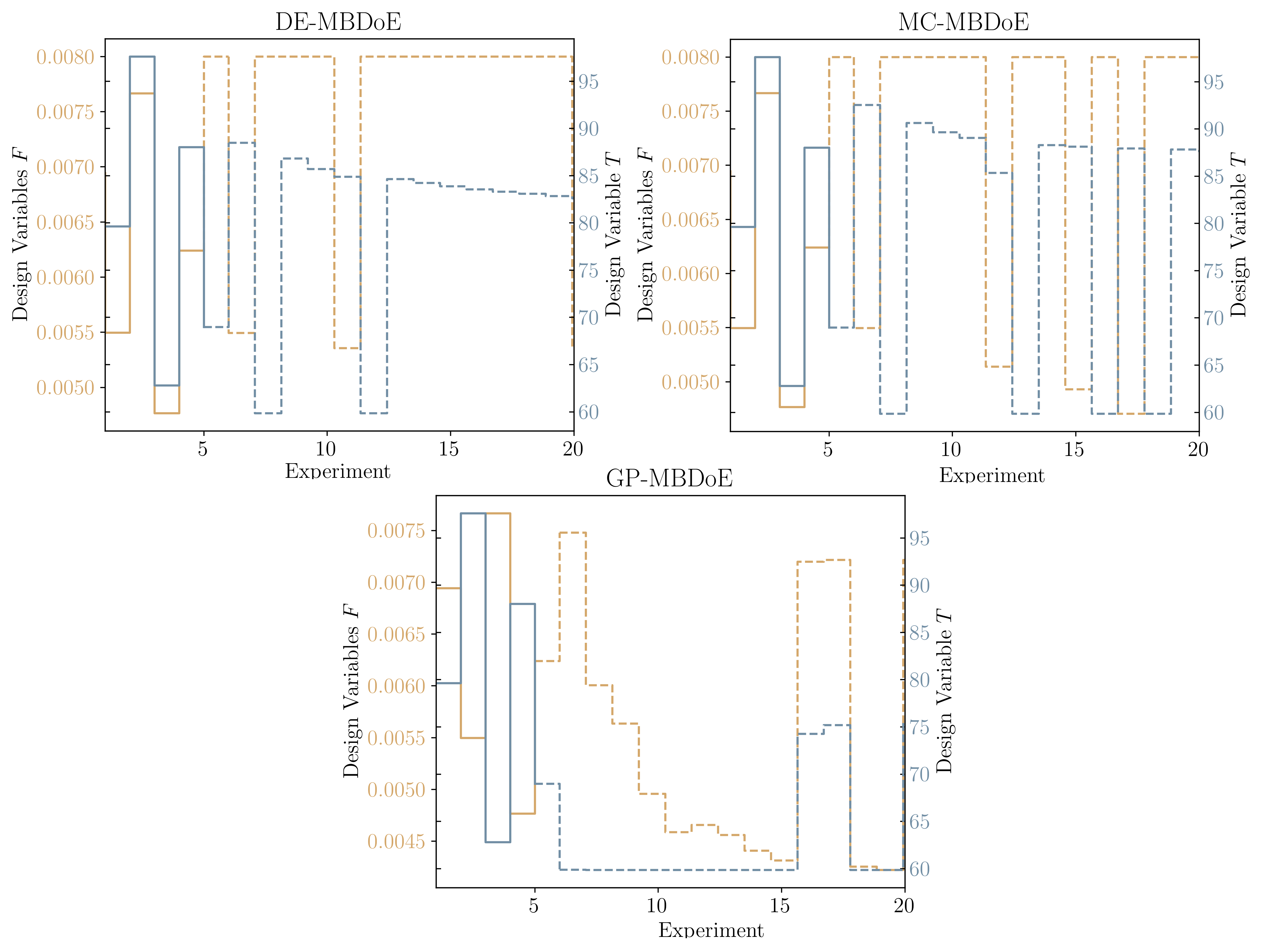}
    \caption{Case Study 2:Computed design variables for the different designs. }
    \label{fig:design_vars_simple}
\end{figure}

It is clear form Fig. \ref{fig:design_vars_simple}, that the 3 frameworks result in 3 fundamentally different strategies. The DE-MBDoE converges to high temperatures and flowrates, and the design variables from MC-MBDoE oscillate. Conversely, the GP-MBDoE appears to focus on smaller values as the number of experiments increases.

\subsection{Case study 2: Design Variables}\label{DV2}
\begin{figure}[H]
    \centering
    \includegraphics[scale=0.2]{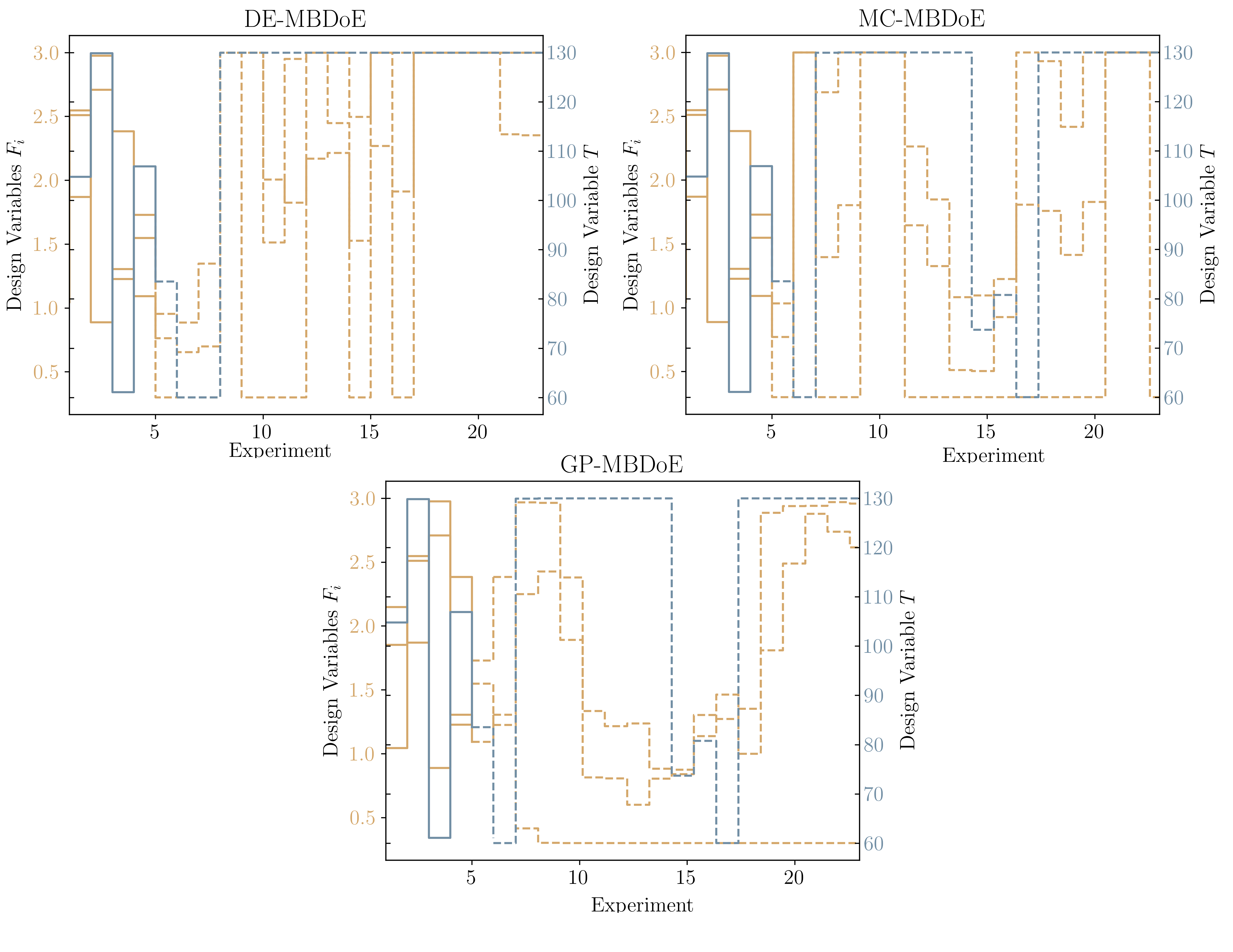}
    \caption{Case Study 2:Computed design variables for the different designs. }
    \label{fig:design_vars}
\end{figure}

The results for the design variables of the second case study are depicted in \ref{fig:design_vars}. All the methods seem to prefer the use of high temperature as the number of experiments increases. Specifically, both MC-MBDoE and GP-MBDoE provide the same designs for the temperature. 
On the other hand, the flowrates are designed differently, where MC-MBDoE results in oscillations and the flowrates of DE-MBDoE converge to the highest value. GP-MBDoE selects to operate in the lowest flowrate for $F_1$ and high flowrates for $F_2$ and $F_3$.
 \end{document}